\documentclass[preprint,12pt]{elsarticle}
\immediate\write18{texcount -inc -sum main.tex > main-wordcount.txt}

\usepackage{graphicx} 
\usepackage{xcolor}
\usepackage{amssymb}
\usepackage{amsmath}

\usepackage{times}
\usepackage{comment}
\usepackage{array, multirow}
  \usepackage[normalem]{ulem}

\usepackage[english]{babel}

\usepackage{graphicx} 
\usepackage{longtable}
\usepackage{hyperref}
\journal{Computer Standards and Interfaces}

\begin{document}

\begin{frontmatter}



\title{ Self-Sovereign Identity and eIDAS 2.0: An Analysis of Control, Privacy, and Legal Implications}


\author{Nacereddine Sitouah} \author{Marco Esposito}\author{Francesco Bruschi} 
\affiliation{organization={Polytechnic University of Milan},
            addressline={ Via Giuseppe Ponzio, 34}, 
            city={Milan},
            postcode={20133}, 
            state={MIlano},
            country={Italy}}



\begin{abstract}

European digital identity initiatives are grounded in regulatory frameworks designed to ensure interoperability and robust, harmonized security standards. The evolution of these frameworks culminates in eIDAS 2.0, whose origins trace back to the Electronic Signatures Directive 1999/93/EC, the first EU-wide legal foundation for the use of electronic signatures in cross-border electronic transactions. 
As technological capabilities advanced, the initial eIDAS 1.0 framework was increasingly criticized for its limitations and lack of comprehensiveness. Emerging decentralized approaches further exposed these shortcomings and introduced the possibility of integrating innovative identity paradigms, such as Self-Sovereign Identity (SSI) models.

\textcolor{black}{In this article, we contribute to the ongoing legal and policy debate on the European Digital Identity Framework by analyzing key provisions of eIDAS 2.0 and its accompanying  recitals, drawing on a systematic literature review guided by defined Research Questions (RQ). This work employs a structured methodological approach that combines descriptive and comparative analysis, systematic gap analysis supported by a defined scoring matrix, and normative analysis to evaluate the compatibility of SSI properties with eIDAS 2.0 regulation, as operationalized via its Architecture and Reference Framework (ARF). Furthermore, we assess  the ARF's guidelines and examine the extent to which it aligns with SSI. The analysis adopts a complementary perspective demonstrating how the regulation can be further developed to better support SSI in the future by identifying existing limitations and potential adoption opportunities within the current legal foundations of the framework.   
}
\end{abstract}




\end{frontmatter}

 
\section{Introduction}

Digital identities today fall into two distinct categories. The first consists of self-generated online identities—such as social media accounts and email addresses—which often carry no legal weight and present underestimated risks \cite{SULLIVAN2018723}. The second category includes formally recognized and regulated identities, such as bank accounts, government-issued electronic IDentification (eID), and business-oriented digital platforms, which are treated as sensitive assets requiring strong protection and informed user management. Furthermore, Internet identification lacks a dedicated layer in Internet protocols, complicating standardization, increasing risks such as identity theft and fraud \cite{preukschat2021self}, and enabling providers of identity platforms to exploit users’ data for profit.

Despite risks, digital identities provide clear benefits. However, although identity providers’ security focuses on external breaches, the necessity of trusting them creates a privacy dilemma.  The current identity management paradigm functions as an infrastructure shaping global internet power dynamics \cite{preukschat2021self}, placing individuals in an unavoidable client role and constraining research on privacy, availability, and data protection. The European Commission (EC) recognized the rapid pace of technological development in the digital era, prompting the adoption of data protection directives such as the General Data Protection Regulation (GDPR) \cite{GDPR}, widely regarded as the world’s strictest privacy and security law. The GDPR is supported by the Data Protection Law Enforcement Directive\footnote{\href{https://eur-lex.europa.eu/eli/dir/2016/680/oj}{Directive (EU) 2016/680} on the protection of natural persons regarding processing of personal data connected with criminal offenses or the execution of criminal penalties, and on the free movement of such data.} ensuring the protection of individuals' personal data whenever it is used by criminal law enforcement authorities for law enforcement purposes (have them be victims, witnesses or suspects of crime). Furthermore, EU member states are required to set up  national data protection authorities to safeguard that citizens' rights and ensure coordinated  enforcement of the GDPR\footnote{\href{https://www.europarl.europa.eu/charter/default_en.htm}{Art.8(3) of the Charter of Fundamental Rights of the EU.}}. 
In addition, the EU opted for a regulation that establishes a unified framework for eIDs and trust services promoting interoperability across state members, enhancing secure electronic interactions and service quality: eIDAS\footnote{\href{http://data.europa.eu/eli/reg/2014/910/oj}{Regulation (EU) No 910/2014} of the European Parliament and of the Council of 23 July 2014 on electronic identification and trust services for electronic transactions in the internal market and repealing Directive 1999/93/EC.} standing for Electronic IDentification, Authentication, and Trust Services, which was amended in 2024 to achieve these objectives, often referred to as eIDAS 2.0.

Modern web and cloud services rely on centralized client-server architectures, introducing risks such as downtime, connectivity issues, and privacy concerns \cite{yeung2023decentralization}; these issues are inherited by  the current Identity Management systems (IDMS). The success of Blockchain networks \cite{hashemi2020cryptocurrency} and Smart Contracts 
opened the door for truly decentralized applications (Dapps) \cite{7}, which was in general made thanks to the emergence of Distributed Leger Technologies (DLTs) \cite{rauchs2018distributed}. These ledgers constitute a consensus record with cryptographic audit trails maintained by a peer-to-peer network and validated by its nodes. Moreover, these technologies have enabled the exploration of new paradigms, such as Self-Sovereign Identity (SSI) models \cite{preukschat2021self}; SSI allows identity holders complete management and control over their digital identities. However, SSI models conflict with eIDAS 2.0  due to differences in some of their objectives.

Current literature on digital identity primarily focuses either on the potential regulatory implications of eIDAS 2.0 or the technological implementation of SSI. Few studies rigorously examine the compatibility of SSI principles with the eIDAS 2.0 Architecture Reference Framework (ARF) \cite{eudi-arf-v280}. 
This work employs a structured  methodological approach  that combines descriptive and comparative analysis, systematic gap analysis and normative analysis to evaluate the compatibility of SSI principles and the eIDAS 2.0 regulation, as operationalized through its ARF. \textcolor{black}{It examines areas of alignment and points of friction, while highlighting how specific architectural and legal choices influence the extent to which SSI concepts can be incorporated into the EUDI framework. }


The theoretical contribution of this work lies in turning a fragmented body of SSI and eIDAS literature into a coherent evaluative lens for legal and technical analysis. Rather than treating SSI as a fixed or universally agreed model, the study consolidates recurring properties from prior work into an assessment structure that can be used to examine regulatory and architectural choices in a more systematic way; in particular: 
\begin{itemize}
    \item We consolidate recurring SSI properties from diverse sources into a coherent evaluation framework.
    \item We translate these properties into criteria for a systematic compatibility assessment.
    \item We compare compliant centralized identity mechanisms with their decentralized counterparts. 
    \item We apply the framework to eIDAS 2.0 and its ARF.
    \item  We identify targeted technical and regulatory adjustments to strengthen and improve SSI-compliant implantation within the European Digital Identity (EUDI) ecosystem.    

\end{itemize}

Thus, the significance of this work lies in clarifying what the compatibility question means for the future design of the European Digital Identity ecosystem with respect to self-sovereignty properties.  \textcolor{black}{ The study details which tensions matter for wallet design, governance, privacy safeguards, interoperability, and compliance choices, thereby offering guidance to policymakersand framework developers of the EUDI system.}

The remainder of the paper is structured as follows:  Section \ref{eIDAS 2.0 history and analysis} provides a background digital identity, SSI and eIDAS regulation. It also presents a literature review on SSI and eIDAS 2.0 compatibility.  Section \ref{chap3} explains the methodology of this study, the approach of a combination of descriptive,  comparative, systematic and normative analysis. Section \ref{Analysis: SSI Property}  provides a detailed assessment of eIDAS 2.0 compatibility with  SSI properties. \textcolor{black}{Section \ref{Findings and Discussion} explores implications and complementary suggestions that should be highlighted for practitioners, researchers and policy makers}. In Section \ref{chap6}, the conclusion offers a  perspective on our analysis and presents an outlook for future research.

\section{Background and Related Work}
\label{eIDAS 2.0 history and analysis}
\subsection{Digital Identity Management}

\begin{figure}[htbp]

\centering 
    \includegraphics[width=0.55\textwidth]{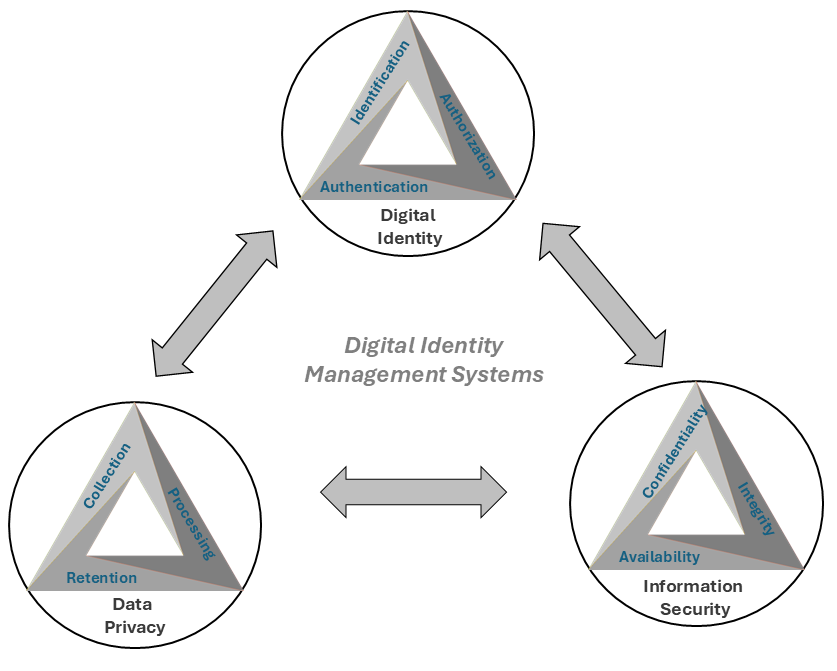}
    \caption{ Identity, Security and Privacy trifecta for an IDMS. }
    \label{id trio}
    \end{figure}

 A Digital Identity or electronic identity (eID) is an electronic reference that uniquely links a person or entity to its digital presence \cite{sedlmeir2021digital,10.1145/3407023.3407026}. Digital identity management can be viewed as a branch of information security, encompassing access control, identification, authentication, and authorization \cite{andress2014basics}. Conversely, from an identity-governance perspective, security and privacy are considered subcomponents of identity management \cite{windley2005digital}.
Identity, security, and privacy form an interdependent trio in digital identity systems. This interdependence is illustrated in figure \ref{id trio}.

\subsection{Digital Identity Models}

Current IDMS mostly adopt centralized or federated architecture, typically implemented through protocols such as OpenID Connect (OIC), Security Assertion Markup Language (SAML), and OAuth \cite{bertino2010identity}. Centralized models rely on a single identity provider for authentication, while federated models enable multiple service providers to share that provider. In a typical OAuth flow, if Emma wants to use David’s application, she is redirected to Carol’s identity service, which authenticates her and vouches for her identity to David. While this simplifies the user experience, it introduces several limitations: the identity provider becomes single point of failure or poses misuse of personal information risks, can potential limit  the range of attributes it can verify, and have the ability to arbitrarily restrict access to services.
 
An emerging alternative model is the so-called "Self-Sovereign Identity" (SSI) \cite{1} or Decentralized Identity, based on identifiers, digital signatures, and credentials. In this model, users are represented by identifiers often linked to cryptographic keys. Anyone can act as an issuer, issuing a signed document (a credential) attesting to something (e.g., that Emma has earned a degree). The credential is then handed over to a holder, who can subsequently present it to a verifier who verifies its origin through the signature. Christopher Allen \cite{1} listed  ten principles that an SSI solution should follow:   Existence, Control, Access, Transparency, Persistence, Portability, Consent, Interoperability, Minimalization and Protection. 
These principles form the foundational framework that has shaped discussions around SSI, but are not universally agreed upon and remain debated. \cite{muhle2018survey, schardong2022self, pava2024self}
 
A fully  decentralized SSI is built on Decentralized Ledger Technologies~(DLTs), which offer transparency, encryption, and true decentralization. A DLT is a system that enables simultaneous access, validation, and updating of records across a peer-to-peer network \cite{8}. Blockchain is one implementation of DLT, allowing participants to track changes, ensuring data reliability and reducing security audits  \cite{7}. 
Blockchain architecture supports the management of SSI by providing a decentralized and tamper-resistant infrastructure for storing and resolving Decentralized Identifiers (DID), and seamlessly validating Verifiable Credentials (VC) \cite{9}. It enables trustworthy verification of identities and credentials without relying on a central authority, since public keys, revocation registries, and identity proofs can be anchored on-chain. 

\subsection{eIDAS 1.0 \& Limitations}

\begin{figure}[h]

\centering 
    \includegraphics[width=1.02\textwidth]{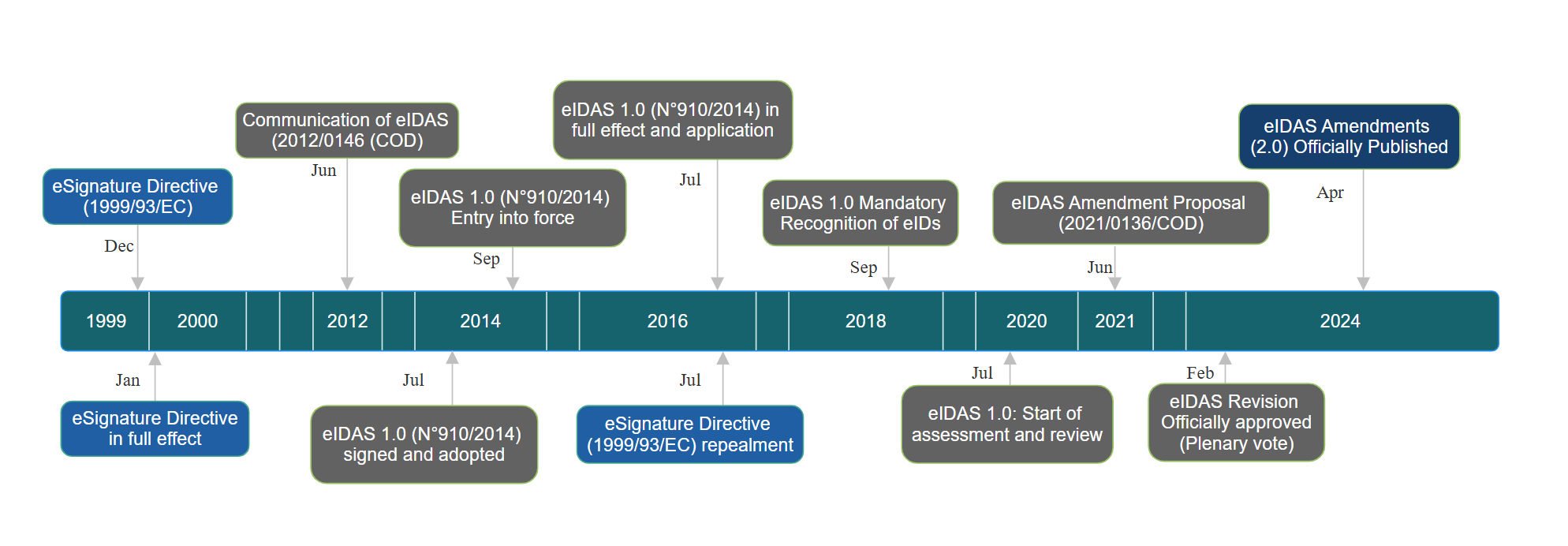}
    \caption{ eIDAS 2.0 Time Line Summary Until Adoption }
    \label{eIDAS timeline}
    \end{figure}

The lack of flexibility and shortcomings of eIDAS 1.0 resulted in only 15 members notifying electronic identity schemes by 2020\footnote{Approximately 6 years after eIDAS 1.0 came into force, while 5 member states notified a scheme as recent as 2023, and Romania  notified their eID scheme (ROeID) in September 2024---\href{https://ec.europa.eu/digital-building-blocks/sites/spaces/EIDCOMMUNITY/pages/48762251/Overview+of+pre-notified+and+notified+eID+schemes+under+eIDAS}{Overview of pre-notified and notified eID schemes under eIDAS}}; while it is 2026, few states\footnote{Namely: Finland, Hungary, Greece and Ireland} have not yet completed the development or/and the testing of their corresponding eID schemes. Relevant with article 49 of eIDAS 1.0, the EC was mandated to review the application of its legislation acts and report back to the parliament by the end of the first semester of 2020.  The report dating to Oct 2020 highlighted areas of weaknesses and improvement and proposed fundamental changes to support missing identification use cases and emerging technologies. 
A voting took place  and concluded with a majority vote for amending regulation No 910/214\footnote{Voting result REGULATION OF THE EUROPEAN PARLIAMENT AND OF THE COUNCIL amending Regulation (EU) No 910/2014 as regards establishing the European Digital Identity Framework Adoption of the legislative act 4016th meeting of the COUNCIL OF THE EUROPEAN UNION (Agriculture and Fisheries) 26 March 2024, Brussels.}. Figure \ref{eIDAS timeline} highlights major events that took part in the adoption of eIDAS 2.0.   

The Level of Assurance (LoA) in eIDAS measures the confidence in the accuracy and reliability of an eID, helping organizations assess the risks of using it in electronic transactions. Based on the ISO 29115 standard\footnote{\href{https://www.iso.org/standard/45138.html}{ISO/IEC 29115:2013: https://www.iso.org/standard/45138.html}}, eIDAS defines three levels of identity assurance: Low\footnote{Single-factor authentication, presenting any identity is accepted for self-registration, and authentication relies on secure methods such as SMS, Smartphone apps, OTPs, etc.} (equivalent to LoA2 of ISO 29115), Substantial\footnote{Two-factor authentication (2FA) is mandatory and verifying identity information is required. } (equivalent to LoA3 of ISO 29115) and High\footnote{It requires the usage of Multi Factor Authentication (MFA), particularly at least two factors; enrollment requires in person verification and registration; while authentication must rely highly secured methods such as biometrics, PKI eID or smart cards, Sim applets, ..etc.} (equivalent to LoA4 of ISO 29115).

Several shortcomings emerged before the revision of eIDAS 1.0, stemming both from the regulation itself and from inconsistencies among Member States’ digital identity frameworks, due to national regulations taking precedence and preventing true cross-border interoperability, despite EU-level peer review of these frameworks. The regulation’s rigidity and complexity also led to unintended exclusions---particularly of private and corporate sectors, which would have required additional national legislation and introduced even more divergence. These integration challenges, combined with the limited availability of notified eIDs, constrained the participation of private service providers and resulted in identity systems unable to support use cases beyond the narrow scope of eIDAS 1.0.
Additionally, Digital identities can support far more applications than traditional identity documents, incorporating attributes, claims, signed attestations, and credentials that make them richer digital structures rather than simple digitized records. However, despite this potential, limited influence from independent and academic researchers has resulted in insufficient interdisciplinary collaboration between legal scholars and computer scientists, slowing progress in adopting SSI \cite{LAATIKAINEN2025102859,satybaldy2024taxonomy}.

The ARF was recommended in 2012 as a development toolbox for the eIDAS regulation and has provided standards, technical specifications, and guidelines. Although it is often referenced for implementation and interoperability, the ARF has no legal force; only eIDAS 2.0 regulation and its implementing acts are legally binding.
The latest ARF\footnote{v2.8.0}  introduces guidance for the European Digital Identity Wallet  (EUDIW), including its core functions and interactions. Unlike the first eIDAS version, eIDAS 2.0 plans to use the ARF as an implementation reference to support the development of EUDIWs, reduce compatibility issues between Member State solutions, and avoid past divergences in notified eIDs. The  EUDIW aims to provide citizens with a secure, reliable digital identity across the EU.

These limitations prompted a major revision of eIDAS 1.0, addressing most of its shortcomings to improve interoperability across the EU market and better integrate private and business sectors.

\subsection{Related Work on SSI and eIDAS 2.0 Compatibility}

\begin{figure}[h]
\centering 
    \includegraphics[width=0.6\textwidth]{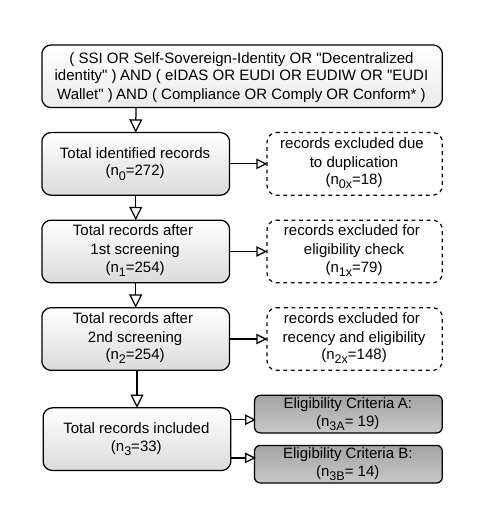}
    \caption{ Existing related research SLR flowchart }
    \label{Existing related research SLR flowchart}
    \end{figure}

In order to obtain a full context related work, we conducted a preliminary Systematic Literature Review (SLR) that follows  the methodology and  guidelines in \cite{kitchenham2007guidelines} and \cite{CARRERARIVERA2022101895} respectively. The guidelines encompass identification, screening for exclusion and eligibility assessment.  We used the search query (( SSI OR Self-Sovereign-Identity OR "Decentralized identity" ) AND ( eIDAS OR EUDI OR EUDIW OR "EUDI Wallet" ) AND ( Compliance OR comply OR Conform*)) across multiple academic databases, including IEEE Xplore, the ACM Digital Library, ScienceDirect, and Scopus. Fig. \ref{Existing related research SLR flowchart} showcases the SLR Flow diagram that outlines the identified results in different stages of the screening. First, we identified  272 records with 18 duplicates, then 148 records were excluded due to format and relevancy\footnote{Exclusion criteria to remove ineligible format (Technical Reports, Proceedings Overview, Extended abstracts, Editorials and Summaries) and records that were submitted before the official final draft of eIDAS 2.0.}, while 79 excluded for not meeting our Inclusion Criteria (IC). The entire screening process considered only documents available before January 2026. Ultimately 33 records remained after full-text analysis; inclusion based on  IC\_A comprising articles that critically evaluate the compatibility of eIDAS 2.0 with SSI or its specific aspects, and IC\_B comprising articles that propose solutions based on stated assumptions or interpretative perspectives on the regulation.  

The findings reveal significant inconsistencies across the literature. Although the issue under consideration is widely acknowledged, no contributions to date explicitly address SSI compatibility following the amendment of eIDAS 2.0. Prior to this amendment, several researchers assumed that Self-Sovereign Identity (SSI) would ultimately be recognized under eIDAS 2.0 \cite{schwalm2022eIDAS,  schwalm2021self, preukschat2021self, pohn2021eid, mattei2024self}. This assumption stemmed from the ambiguity of both eIDAS 1.0 and early eIDAS 2.0 drafts, which had many aspects open to interpretation, as well as from the fact that the ARF at the time explicitly referenced SSI and Verifiable Credentials. This continues to generate confusion, particularly due to the divergent and sometimes unrelated definitions of credentials in the OIC and W3C ecosystems, which is discussed in depth in \cite{biedermann2024systematisation}. Moreover, Such assumption and prediction---especially those presuming conformity---carry notable repercussions, which are highlighted in the following.

Alvarez et al. \cite{alvarez2025privacy} provide an analysis of the EUDIW by employing a list of qualitative privacy risk assessment methods in order to map and evaluate information flow of attestation issuance and presentation, then identify privacy risks such as linkability, identifiability and personal information disclosure. Indicating that while the EUDIW is somewhat a next step for SSI, it still does not even fulfill all its privacy commitments. They suggest to focus research and adopt standards with advanced Privacy Enhancing Techniques (PETs) such as ZK proofs, network layer anonymization and anonymous credentials for EUDIW. Similarly, \cite{podda2025impact} analyzes the legal compatibility of PETs with eIDAS, emphasizing that regulatory bodies should enforce their use, as conventional cryptographic methods---although recognized as compliant---may conflict with GDPR requirements. Therefore, the regulatory challenges associated with PETs must be addressed. Kutylowski et al. \cite{kutylowski2025pseudonymization} further elaborate on the importance of PETs using the cases of whistleblowers---reporters of unethical, illegal or harmful activities. According to authors assessment, the ARF suggests privacy cryptographic techniques that fail to meet privacy requirements outlined in eIDAS 2.0, since those pseudonymisation methods can be traceable by privileged bodies, which can result in unjust retaliation. Further, privacy is a major concern that is repeatedly stated in the literature \cite{babel2025self, althabhawi2025mutual, sakka2024blockchain, herbke2024decentralized}  

Moser et al. \cite{moser2025privacy} propose a bridging solution that leverages ZK proofs to achieve eIDAS-compliant, GDPR-preserving digital signatures for blockchain-based SSI systems. CONSENTIS \cite{kyriakoulis2025consentis} is an SSI-based consent management solution proposed for integration with the EUDIW, incorporating DIDs, VCs, and a consent policy manager operated alongside  smart contracts. There are also other digital identity bridges in \cite{biedermann2025aggregating, lepore2024aligning} that incorporates Web3 identities to OIC protocols and EUDIW credentials.

Scalable cloud-based solutions are also considered a viable approach to improving the availability of identity wallets, such as cryptoscape \cite{alfughi2025cryptoscape}. Bukhari et al. \cite{bukhari2024defining} propose libraries that create a unified user-selected signature API to improve the interoperability of EUDIW. In \cite{11260514}, authors focus on multi-authority credentials mandated by eIDAS 2.0, highlighting that a privacy preserving system that supports credential aggregation requires decentralization, unlikable authentication and scalable verification. They propose OSDISC, a credential system using constant-size authentication tokens, ZK proofs, and cryptographic commitments to enable independent credential aggregation.

 Pohn et al. \cite{weigl2025governance} applied an institutional isomorphism in order to investigate the regulation's governance trends. Although the study showcases a shift in identity governance, control and power in data governance creates a conflict of interest. The major identified issue is the need for governance frameworks that keep regulatory bodies independent from the bodies they authorize and overseas. Thus, clear and robust independence requirements are necessary to safeguard privacy, accountability and ensure trust. Another problem with the EUDIW certification is the potential diverse interpretations of standards and security measures that lead to more divergence and fragmentation \cite{loutocky2025certification}. Furthermore, a social entrepreneurship study \cite{vano2025social} shows that eIDAS 2.0 disproportionately burdens smaller social enterprises, particularly because of its reliance on costly blockchain technologies. The authors emphasize that interoperability should be grounded in inclusive equity and technological neutrality, not solely regulatory compliance.

The literature shows that a wide range of DLT technologies has been extensively studied, highlighting an urgent need for legal scholars and policy makers to pay attention and engage more with technology developers \cite{Singh202527}. As the implications for interoperability, accessibility, and transparency are substantial and cannot be overlooked. There is also emphasize that EU-based projects such as EBSI could play a significant role in decentralization by introducing decentralized PKI as infrastructure for the EUDIW, thereby enhancing trust \cite{alamillo2025role}. While many  still rely on academic records that we consider inaccurate or out-dated after the official amendment, some under the believe that eIDAS 2.0 a shift towards decentralized data storage \cite{loutocky2025certification, vano2025social, bochnia2024long}, decentralized authentication \cite{boi2025user, gosslbauer2025towards, bochnia2024self}, SSI adoption \cite{kyriakoulis2025consentis, ibor2025considerations, mazzocca2025survey, arshad2025web3, ahmed2025distributed, petrlic2024ssi, morales2024towards} or PET-based security requirements \cite{ramos2025regulatory}.

Table  \ref{SSI Principles and Representative Related Work} presents an overview of SSI principles covered in the SLR, offering a perspective on their alignment with eIDAS 2.0.

\begin{table}[t]
\centering

\label{tab:ssi-survey}
\scriptsize
\begin{tabular}{p{1cm} p{1.78cm} p{1.6cm} p{1.7cm} p{1.5cm} p{1cm} p{1.75cm} }
\hline
\textbf{Relevant record} & Privacy \& minimal Disclosure & Decentralization & Verifiability \& Authenticity & Interoperability & Security & Accessibility \& Transparency \\
\hline

\cite{alvarez2025privacy, podda2025impact, kyriakoulis2025consentis, kutylowski2025pseudonymization, althabhawi2025mutual, herbke2024decentralized} & $\checkmark$ & -& -&- & -  &-\\ 
\cite{11260514, sakka2024blockchain}  & $\checkmark$ & $\checkmark$ & $\checkmark$ &- &- &- \\
\cite{moser2025privacy} & $\checkmark$ & $\checkmark$ & - &$\checkmark$ &- &- \\

\cite{babel2025self} & $\checkmark$ & - & - &$\checkmark$ &- &- \\
 \cite{weigl2025governance}  & $\checkmark$ & $\checkmark$ & - & $\checkmark$ & $\checkmark$ & - \\
 \cite{vano2025social} &- &- & -&  $\checkmark$ & -&  $\checkmark$\\

  \cite{Singh202527} & - & - & - & $\checkmark$  & - &$\checkmark$ \\

  \cite{loutocky2025certification} & - & -& - & $\checkmark$  & $\checkmark$ & - \\

\cite{biedermann2025aggregating}  & - & $\checkmark$  & - &$\checkmark$  & -& -\\

 \cite{alfughi2025cryptoscape} & -& -&- & -&$\checkmark$ & $\checkmark$\\
\cite{alamillo2025role, lepore2024aligning} & -& $\checkmark$&- & -& - & -\\
\cite{bukhari2024defining} & -& -&- &$\checkmark$& - & -\\
\hline
\end{tabular}
\caption{Comparative overview of how the selected literature covers key SSI principles relevant to the compatibility debate around eIDAS 2.0. A checkmark indicates substantive and critical evaluation  of the compatibility with  the corresponding principle.}

\label{SSI Principles and Representative Related Work}
\end{table}

\section{Methodology}
\label{chap3}
\subsection{Research Objectives}
This study adopts a \textcolor{black}{normative-analytical framework } to evaluate the compatibility of SSI with the eIDAS 2.0 regulatory framework. The primary objective is to extend beyond descriptive reporting and establish a rigorous research foundation that systematically identifies tensions, gaps, and \textcolor{black}{complementary recommendations to further align the EUDI framework with SSI principles}.
\subsection{Research Questions}
 This study is guided by the following research questions:
\begin{enumerate}
\item[RQ1]: To what extent are  SSI principles compatible with the eIDAS 2.0 Architecture Reference Framework?
\item[RQ2]: What technical or regulatory modifications could improve alignment with GDPR and ARF requirements while preserving SSI principles?
\item[RQ3]: Which  challenges arise specifically from misalignments between DLTs and existing regulations?
\end{enumerate}
\subsection{Analysis Approach and Assessment Criteria }
SSI is frequently discussed in relation to Allen’s ten principles \cite{1}, which are commonly used as a reference framework for achieving user control over digital identities. However, there remains no clear consensus on which criteria definitively characterize a fully compliant SSI system. In \cite{9858139}, authors discussed this issue and investigated the literature to capture overlapping and missing properties of a true sovereign IDMS. This evaluation study draws from five distinct SSI property frameworks \cite{1,toth2019self,ferdous2019search,stokkink2018deployment,2}, identifying similarities and differences and combining overlapping features into unified properties. We use the resulting properties, slightly altered to assess how eIDAS 2.0 and its ARF align with SSI. Tab.\ref{properties_categl} illustrates SSI properties classification reported in \cite{9858139}. Our analysis is structured according to these SSI properties, where each property is assessed against eIDAS 2.0. This property-by-property compliance study allows for a detailed examination of alignment, incompatibilities, and areas requiring \textcolor{black}{regulatory revisions or complementary functionalities to achieve compliance}.
   \begin{footnotesize}
    \begin{longtable}{|p{2.1cm}|p{5cm}|p{5.2cm}|}
    \caption{Categorization of SSI properties according to \cite{9858139}.}
    \label{properties_categl}
       \\ \hline
        Category & Definition & List of properties \\ \hline
          \multirow{3}{*}{Controllability }  & \multirow{3}{=}{Properties that allow entities to gain and  maintain control over their identities.}  & 1. Existence and Representation \\ & & 2. Decentralization and Autonomy \\ & & 3. Ownership and Control\\ 
          
          \hline
 \multirow{3}{*}{Privacy} & \multirow{3}{=}{Properties that allow entities to preserver privacy while interacting with other entities.} & 1. Privacy and Minimal Disclosure\\ & & 2. Single Source and Consent \\ & & \\\hline
         \multirow{4}{*}{Security} & \multirow{4}{=}{Properties that maintain data security, authentication and authorization of entities during interactions} & 1. Security \\ & & 2. Protection \\ & & 3. Verifiability and Authenticity \\ & & \\ \hline
         \multirow{5}{*}{Usability} & \multirow{5}{=}{Properties that affects users feeling when they experience the self-sovereign digital identity system and  services.}  & 1. Accessibility and Availability \\ & & 2. Recoverability \\ & & 3. Usability and User experience \\ & & \\ & & \\\hline
         \multirow{7}{=}{Adoption and sustainability} & \multirow{6}{=}{Properties that provide acceptance of self-sovereign identity models.} & 1. Transparency \\ & & 2. Persistence \\ & & 3. Interoperability \\ & & 4. Portability \\ & & 5. Compatibility with legacy systems \\ & & 6. Usability and User Experience \\ & &  7. Cost\\ \hline

\end{longtable}
\end{footnotesize}


The analysis of each relevant SSI property or cluster proceeds in the following stages:  
\begin{enumerate}
 
    \item \textbf{Definition:} a concise description of the SSI property, or cluster of interdependent properties.  
    \item \textbf{Relevance to Digital Identity}: an account of how the property supports secure user-controlled identity systems, with examples of systems that fully implement this property and those that fail to do so.
    \item \textbf{Compatibility assessment with eIDAS 2.0 ARF: }an evaluation of alignment or divergence with the regulatory requirements.
    \item \textbf{Challenges \& gaps:} an identification of legal, technical, or operational tensions.
    \item \textbf{Recommended improvements:} a guidance on improving alignment, compliance, and interoperability.
\end{enumerate} By combining normative legal analysis with technical evaluation of standards, this methodology ensures that the study produces research-based findings rather than merely descriptive observations.
We assess compatibility through a qualitative \textcolor{black}{four level scale consisting of Compatible, Non-compatible, Partially-compatible: Architecturally limited, and partially-compatible: Optional coverage } (see \ref{app:assessment-rubric-traceability}). To reduce interpretive bias, each judgment is grounded in the text of the regulation, the relevant ARF mechanisms, and the supporting literature. Where the framework leaves room for national implementation choices, we describe the result as conditional rather than complete.

 \section{Assessing eIDAS 2.0 Against SSI Criteria}\label{Analysis: SSI Property}
\begin{figure}[h]

\centering 
    \includegraphics[width=0.95\textwidth]{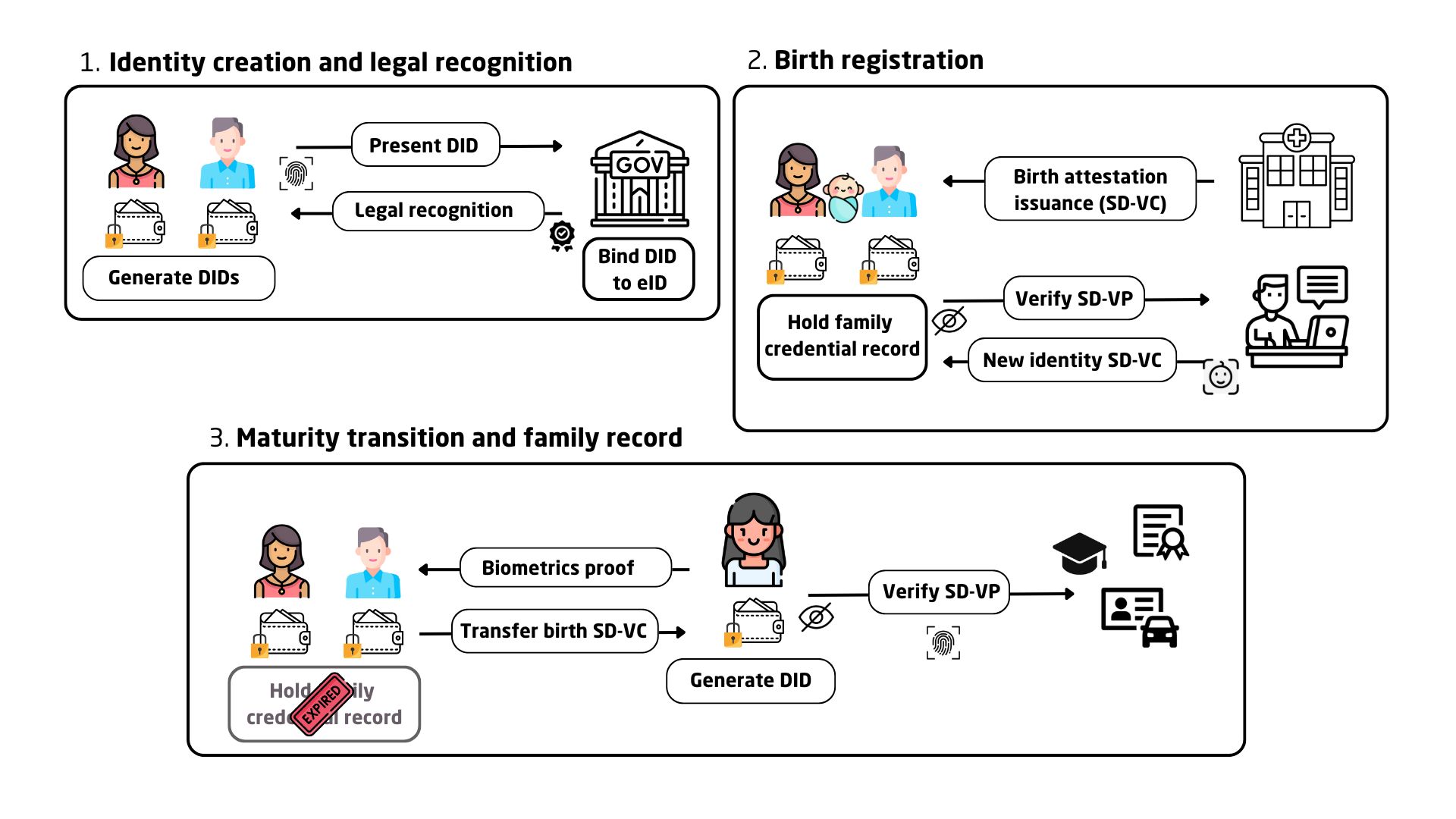}
    \caption{SSI-based identity lifecycle showing: (1) DID generation and legal recognition process, (2) birth registration using SD-VC (Selective Disclosure Verifiable Credentials) and SD-VP (Selective Disclosure Verifiable Presentations), and (3) maturity transition with credential transfer.}
    \label{fully ssi}
    \end{figure}

We use a real-life scenario to illustrate how identities are issued and managed in an SSI system and how key properties are ensured. In such a system, control shifts to citizens while preserving trust with verification authorities. Each user---such as Emma and David---holds a sovereign digital identity wallet with cryptographic and biometric security, not governed by any central authority. They generate their own DIDs and later link them to government records for legal recognition, without surrendering control of the identifiers. All administrative interactions occur through the user’s wallet using selective disclosure, allowing citizens to decide exactly which information is shared.     When Emma and David have a child (Sofia), a trusted medical authority issues a birth attestation with minimal required data, stored securely by the authority but delivered directly to the parents’ wallets. To register Sofia, they share only the necessary parts of this attestation with municipal authorities, who then issue an official identity credential also stored in the parents’ wallets. Upon reaching maturity, Sofia creates her own wallet and DIDs, and after biometric verification, her credentials are transferred to her control, with previous versions invalidated. From then on---whether applying for university, obtaining a driver’s license, or receiving professional certificates---she uses her wallet to prove required attributes with selective disclosure, minimizing unnecessary exposure of personal information, as illustrated in Figure. \ref{fully ssi}.

 \subsection{Existence and Representation}

    \subsubsection{Definition } This refers to the individual's ability to create as many digital identities as they need without reliance on third parties for issuance. The ability to verify or present identity is independent of any external authority, and exercising such identity does not depend on the whim or cooperation of anyone. 

     \subsubsection{Relevance to Digital Identity}
    Existence and Representation as a property is a foundational SSI principle, in contrast to most centralized systems where an individual digital presence is not independent as it is meant to be. State-provided eID schemes\footnote{Inc. notified eIDs such as CIE, SPID, German eID, etc.}, \textcolor{black}{typically built on centralized OAuth-based protocols, do not satisfy this property, as they prevent identities from existing independently of the issuing authority}; users may hold only the identities they are authorized to possess, mostly limited to one per scheme. Access to these identity services can also be restricted or denied at the provider’s discretion \cite{sharif2022eIDAS}. The same applies to electronic mailing services\footnote{Such as Gmail, Outlook, Yahoo, etc.} and social media accounts: although users may freely create multiple accounts, these identities depend entirely on providers whose business models monetize user data. Providers retain unilateral control and may change terms, restrict access, or terminate accounts at any time \cite{greschbach2012devil}.

 There are also systems that meet this property’s requirements. A key example is traditional graphical signature–based identity, where a user signs a document with a unique handwritten mark that only they can reproduce. Others can verify whether a document was signed by the same individual by comparing signatures, and users are free to create distinctive marks that are difficult or impossible for others to imitate, assuming signatures cannot be easily counterfeited \cite{bhavani2024multi}. Mobile SIM cards also illustrate a partially compliant system. A user can possess a pseudo-digital identity object issued by a Mobile Network Operator (MNO)  the user with control over a unique identifier\footnote{such as a phone number or IMSI} \cite{koshy2018evolution}. As long as government identification is not required, a person can freely possess and control his SIM card\footnote{For this property to hold, neither the MNO nor the user’s government should require official national identification to obtain a SIM card; the MNO must also refrain from censoring communication or blocking access through IMSI blacklisting or invalidating authentication data. When national identification is required, the compliance of the SIM-based identity depends entirely on whether the identification method used during acquisition satisfies the existence property; if it does, the SIM card identity inherits that compliance, and if not, it likewise fails to comply.}.

     \subsubsection{Compatibility Assessment} 
     
     The EUDIW and its associated services are not considered   valid unless the user obtains a Person Identification Data (PID) from a designated provider that meets the required high LoA.
     According to the ARF, these providers are the same organizations that currently issue official identity documents and  electronic identity means,  which can also be EUDIW providers, because art 7(d)\footnotemark{}\footnotetext{Letter d of article 7 from \href{http://data.europa.eu/eli/reg/2024/1183/oj}{Regulation (EU) 2024/1183}: "the \textbf{notifying Member State} ensures that the \textbf{person identification data} uniquely representing the person in question is attributed, in accordance ... to the natural or legal person referred to in ... at the time the electronic identification means under that scheme is issued; "} reserves the right to take into account only PIDs that were assigned by member states (or bodies accredited and designated by them) and 
 to limit each user to one unique PID\footnotemark{}\footnotetext{Recital (19) from \href{http://data.europa.eu/eli/reg/2024/1183/oj}{Regulation (EU) 2024/1183}: "... Only Member States’ competent authorities can provide a high level of confidence in establishing the identity of a person and therefore provide assurance that the person claiming ... It is therefore necessary for the provision of European Digital Identity Wallets to rely on the legal identity of Union citizens ... "}. \textcolor{black}{Nonetheless, the ARF recommends an operational wallet state that enables  users to  access and manage EAAs, namely credentials issued by non government or certified entities. This approach offers users greater flexibility in acquiring credentials, albeit with reduced legal assurance, as eIDAS formally recognizes only PID-based electronic identification. It may also signal a gradual openness toward the use of reputational or decentralized credentials by citizens, which can complement centralized certificates, while remaining within the bounds of the regulation.}
 
 \textcolor{black}{A potential risk} resides with EUDIW instances and their providers, a EUDIW provider holds the responsibility of assuring the overall validity and security of each instance, this is conceptualized in the ARF in a WUA to verify that the provider still trusts this instance. This centralized approach proposed in the ARF indeed presents effective risk management,  in accordance with Art. 5e(1-2)\footnotemark{}\footnotetext{Letter e of article 5 from \href{http://data.europa.eu/eli/reg/2024/1183/oj}{Regulation (EU) 2024/1183}: "Point 1: '... Where justified by the severity of the security breach or compromise ... the Member State shall \textbf{withdraw European Digital Identity Wallets without undue delay}....' 
Point 2: 'If compromised wallet not remided within three months of suspension, it should be withdrawn and revoked validity.'"}, such as minimizing damage when instances are compromised rather than disabling the whole wallet service app. However, not only the management of all these instances is a difficult task and might cause scalability issues, misusage of the ability to selectively deny access to an EUDIW without decentralization and public transparency \textcolor{black}{marks an architectural conformity gap with} the principle of existence and the right of representation     
\footnotemark{}\footnotetext{Point 15 of article 5a from \href{http://data.europa.eu/eli/reg/2024/1183/oj}{Regulation (EU) 2024/1183}: "...Access to public and private services, access to the labour market and freedom to conduct business shall not in any way be restricted or made disadvantageous to natural or legal persons that do not use European Digital Identity Wallets ..."} \footnotemark{}\footnotetext{Recital (15) from \href{http://data.europa.eu/eli/reg/2024/1183/oj}{Regulation (EU) 2024/1183}: "... Member States should not, directly or indirectly, \textbf{limit access} to public or private services to natural or legal persons not opting to use European Digital Identity Wallets and should make \textbf{available appropriate alternative solutions}. ... "}. This means that, under the ARF, both Emma and David must obtain a valid PID from the competent authority in their Member State to access and use the EUDIW, which will likely be linked to a notified eID in accordance with Recital (29)\footnotemark{}\footnotetext{Recital (29) from \href{http://data.europa.eu/eli/reg/2024/1183/oj}{Regulation (EU) 2024/1183}: "The objective of this Regulation is to provide the user with a fully mobile ... As a transitional measure until the availability of certified tamper-proof solutions, such as secure elements within the users’ devices ... European Digital Identity Wallets should be able to rely upon certified external secure elements ... or upon notified electronic identification means at assurance level high ..."} if they have LoA high, or other transitional external secure methods.

\subsubsection{Challenges \& gaps}
\begin{itemize}
    \item The legal requirement of \textcolor{black}{exclusively} unique government-issued PIDs \textcolor{black}{creates a centricity that architecturally constrains the potential of attribute-based credentials}.
    \item The dependence on designated EUDIW instances and endpoints to obtain WUA status \textcolor{black}{partially} undermines the sovereignty of digital existence.    
\end{itemize}
\subsubsection{Recommended improvements}
To \textcolor{black}{improve structural}  access restrictions and support continuous digital representation:
\begin{itemize}
    \item Allow natural persons to access and operate their wallet instances as long as both the wallet and the hosting device remain secure, \textcolor{black}{potentially user-controlled self-custody EUDI wallets or emerging cloud-based wallets \cite{sabanic2025confidential, zhou2025dstack}}.
    \item Support the creation of anonymous and pseudonymous identifiers by users, enabling individuals to maintain digital identities even without possessing a valid PID.
    \item Ensure that the validity state of an EUDIW does not depend on holding a valid PID; instead, allow PID-dependent services to verify PID  only when required, while other services may rely solely on other identifiers.
    \item \textcolor{black}{Recommend, within the exiting regulatory ARF,}   the use of autonomously generated DIDs backed in the Wallet Secure Cryptographic Device (WSCD) or Wallet Secure Cryptographic Application (WSCA), ensuring independent self-representation.
    
    \item \textcolor{black}{Recommend}  the binding of user-generated DIDs to official eIDs and other credentials.

    \item \textcolor{black}{Recommend}   legal persons \footnote{A Legal person varies according to each government laws (registration in some countries is required as soon as the the business is running regardless of its state, while in others it dependents on the type of the business, its size or its annual estimated turnover \cite{vermeulen2012liability}). } to  create multiple DIDs while relying VCs for secure authentication, authorization, and recognition of legitimacy in transactions, recognizing that identifiers alone are not sufficient for legal verification. 
 
\end{itemize}
 
      \vspace{-0.3cm}
 
 \subsection{Decentralization, Transparency and Autonomy} 
 
        \subsubsection{Definition } The requirement that users maintain self-rule and management of  personal identity data independently, autonomous decision control regarding data disclosure, storage, update, deletion and interactions purpose and frequency. Autonomy also refers to the ability to control sharing  and access to official credentials.   \\
        Decentralization\footnote{ While some SSI definitions demand complete decentralization with no central points of control, or rely solely on decentralized infrastructures for identity registration (such as DLTs), we adopt the definition in which decentralization specifically ensures that verification processes occur without intermediaries, even when credential issuance may still rely on centralized systems  \cite{hoffman2020toward, alizadeh2022comparative}. This interpretation provides a practical balance: it preserves user autonomy and minimizes reliance on central authorities during verification, without imposing unrealistic constraints on real-world identity registration mechanisms.} in this context describes the absence of central authorities during identification, authentication, and authorization processes, aligning with the autonomy requirements stated above and must be accompanied by sufficient transparency~---~meaning that identity processing is understandable to users, based on open and architecture-independent methods, and gives all persons full insight into their data and related interactions.
     \subsubsection{Relevance to Digital Identity}
     
      Autonomy and decentralization are interdependent properties of SSI that require  adequate technological and governance mechanisms. IDMSs that rely on centralized servers to store identification data such as PKI-based systems, password-based login, and other server-dependent mechanisms like Kerberos, OIC, SAML do not meet this property's requirements \cite{10.1145/3407023.3407026, cao2010survey}. The same limitations apply to identity cards (physical or electronic), as users rely on issuers to access  identity services, and they cannot manage or control access to  attributes or attestations embedded with their identity. While verification requires relying on the issuer to confirm an identity’s legitimacy, and verifiers can access identity data without consent or transparent trace, undermining the user’s ability to control its distribution. 

     Although mainstream IDMSs are centralized in nature, they sometime deploy mechanism that comply with decentralization, such as   Time-based One-Time Password (TOTP) authentication,  which requires both parties to pre-establish shared parameters---a Hash-based Message Authentication Code (HMAC) OTP algorithm, a start time, and a time interval--to compute synchronized one-time values. The authenticatee generates a TOTP and sends it to the authenticator, who verifies it by comparing it to a locally generated value. This enables autonomous authentication, as users can independently generate their secrets\footnote{Third-party services that store or manage HMAC-TOTP parameters must be avoided, ensuring that shared secrets and generated values remain under the direct control of the two parties. A third party may assist in the initial synchronization only if both sides can independently establish and retain full control over the pre-shared parameters thereafter}. In contrast, SSI-based solutions achieve decentralization and autonomy through blockchain networks and self-created DIDs: users generate DIDs from their blockchain key pairs, validators confirm ownership, and the resulting DID is anchored transparently on-chain, enabling issuers and verifiers to sign and validate credentials without intermediaries.

     \subsubsection{Compatibility Assessment} 

    The regulation \textcolor{black}{does not prohibit decentralized technologies per se, } \textcolor{black}{however, the chosen architecture for the EUDI framework impacts this property.} As legal and natural person are set to be largely dependent on providers for the verification of issued credentials in their wallet, the ARF suggests the usage of OIC protocols leveraging Verifiable credentials (OID4VC, OID4VCI, OID4VP) that rely on trust anchors\footnote{VC in W3C \cite{9} are not to be confused with OID4VC \cite{bochnia2024self}, as SSI support the usage of  OID credentials but not the other way around.}. Although the ARF does not require trust anchors to use traditional PKI, OIC specifications and current implementations rely on centralized PKI models, which constrain the decentralization achievable under eIDAS 2.0. \textcolor{black}{Nonetheless, the adoption of attribute-based verifiable credential represents a significant step towards more flexible and powerful identity and credential frameworks, particularly when compared to existing models based on OAuth 2.0 and single sign-on protocols}. 
    
    Wallet Secure Cryptographic Devices (WSCDs) are intended to give users autonomy in generating credential requests and presentations, but most smartphones lack local WSCDs or OS-integrated options. As a result, users must rely on remote WSCDs or Wallet Secure Cryptographic Applications (WSCAs) to maintain that autonomy\footnote{WSCD refer to hardware-based security elements, such as Secure Elements (SE), Trusted Platform Modules (TPM), or Secure Enclaves, designed to securely store and process cryptographic credentials within a user-controlled environment \cite{ashraf2020analytical}. WSCA are software-based alternatives that implement similar security functions through secure execution environments or remote cryptographic services, ensuring credential integrity and controlled key usage \cite{ebadi2025secure}. }. EU members are encouraged\footnotemark{}\footnotetext{Recital (36) from \href{http://data.europa.eu/eli/reg/2024/1183/oj}{Regulation (EU) 2024/1183}: "... To ensure that the European Digital Identity Framework is open to innovation, technological development and future-proof, Member States are encouraged, jointly, to set up sandboxes to test innovative solutions in a controlled and secure environment ... That environment should foster the inclusion of SMEs, start-ups and individual innovators and researchers, as well as relevant industry stakeholders. ... "} to collaborate with non-governmental entities to promote fairness and reduce centralization, supported by making solution source code publicly available\footnotemark{}\footnotetext{Recital (33) from \href{http://data.europa.eu/eli/reg/2024/1183/oj}{Regulation (EU) 2024/1183}: "... Member States should disclose the source code of the user application software components of European Digital Identity Wallets, including those that are related to processing of personal data and data of legal persons. The publication of this source code under an open-source license should enable society...This would increase users’ trust in the ecosystem and contribute to the security of European Digital Identity Wallets... "}. Recital (19)\footnotemark{}\footnotetext{Recital (19) from \href{http://data.europa.eu/eli/reg/2024/1183/oj}{Regulation (EU) 2024/1183}: "... European Digital Identity Wallets should also allow users to create and use qualified electronic signatures and seals which are accepted across the Union. ... natural persons should be able to use it to sign with qualified electronic signatures, by default and free of charge, without having to go through any additional administrative procedures. Users should be able to sign or seal self-claimed assertions or attributes. ... Reliance on the legal identity should not hinder ... users to access services under a pseudonym, where there is no legal requirement for legal identity for authentication. ..."} affirms users’ autonomy to create self-claimed assertions and use pseudonyms when legal identification is not required. These assertions and pseudonyms must not serve as inputs to electronic identification means, as PIDs are explicitly designated as the proper authentication data unit.  

 eIDAS 2.0 does not reject decentralization via DLTs, adopting a neutral\footnotemark{}\footnotetext{Recital ( 68) from \href{http://data.europa.eu/eli/reg/2024/1183/oj}{Regulation (EU) 2024/1183}: "...To ensure legal certainty and promote innovation, a Union-wide legal framework that provides for the cross-border recognition of trust services for the recording of data in electronic ledgers should be established. This should sufficiently prevent the same digital asset from being copied and sold more than once to different parties. The process of creating and updating an electronic ledger depends on the type of ledger used, namely whether it is centralized or distributed. This Regulation should ensure technological neutrality, namely neither favouring, nor discriminating against, any technology used to implement the new trust service for electronic ledgers. } stance while treating them mainly as record-layer alternatives. Decentralization without DLTs or public networks can still meet the principle, but often makes transparency and privacy difficult or impossible to verify---issues addressed later under the relevant privacy property. Transparency on the other hand is well encouraged, as opaqued methods are prohibited\footnotemark[43]
\footnotemark{}\footnotetext{Recital (13) from \href{http://data.europa.eu/eli/reg/2024/1183/oj}{Regulation (EU) 2024/1183}: "European Digital Identity Wallets should have the function of a common dashboard embedded into the design, in order to ensure a higher degree of \textbf{transparency, privacy and control ...}"}
 \footnotemark{}\footnotetext{Recital (56) from \href{http://data.europa.eu/eli/reg/2024/1183/oj}{Regulation (EU) 2024/1183}: "...Any request by the relying party for information from the user of a European Digital Identity Wallet should be necessary for, and proportionate to, the intended use in a given case, should be in line with the principle of data minimisation and  \textbf{should ensure transparency} as regards which data is shared and for what purposes... . "}.  The regulation treats EUDIW transparency and provider accountability as essential for building social trust and acceptance. The ARF’s privacy-by-design approach supports this through strong encryption, anonymization, and user consent.  \textcolor{black}{However, existing transparency challenges persist given the fact that systems based on provider-centric, centralized OIC federations inherently concentrate control over identity flows and transaction logs within a set of certified intermediaries. Such approach contains the ability of users to independently verify how authentication events, identity attribute disclosures and consent decisions are processed. }

\subsubsection{Challenges \& gaps}

 \begin{itemize}
     \item The OIC architecture inherently imposes a centralized mechanism.
     \item The absence of legal recognition and regulation for DLTs, particularly blockchain networks. 
 \end{itemize}

 \subsubsection{Recommended improvements}

\begin{itemize}
   \item  \textcolor{black}{Recommend}    users to autonomously generate DIDs\footnote{ The DID is added to the peer-to-peer network through a decentralized consensus mechanism, and is under the sole controller of its owner using  its cryptographic key pair \cite{9}.}, ensuring they can create true pseudonymous identifiers that can be associated with EUDI credentials (PIDs, (Qualified)-Electronic Attestation of Attributes (Q)-EAAs, etc) before or after issuance.

 \item Provide a DLT network co-created, governed and controlled by government authorities, private companies and industrial unions (such as QuarkID \cite{quarkid2022whitepaper, sitouah2024methodologies, cambarieri2024explorando} as a decentralized infrastructure hosting DIDs with VCs; enabling privacy-preserving direct and transparent verification of credentials.

\item Ensure EUDIW activation can be performed with a DID bound to a notified eID, enabling users independent private authentication.

\item Alternatively, \textcolor{black}{recommend}  credential status verification \cite{sitouah2024untraceable, fang2020verifiable, khongbantabam2025hierarchical} methods that eliminate the centralized nature of PKI-based Credential Revocation list (CRL) mechanism, coupled with strong transparent user-centric consent system such as Data track of the project “PRIME – Privacy and 
Identity Management for Europe  \cite{hansen2007transparency}\cite{10.1007/978-3-642-34210-3_16}).

\end{itemize} 
 
  \subsection{Ownership. Single Source, control and consent} 
        \subsubsection{Definition }   This property requires that users hold full authority over their digital identities and all associated personal data---including credentials, identifiers, and keys---and act as the single source of truth for that information. Users must be able to manage, share, or delegate access to their identity data, whether self-generated or issued by trusted parties, and no third party may use or exchange this data without their explicit, informed, and revocable consent. True ownership implies continuous control over identity and its disclosures, ensuring that all interactions occur with the user’s knowledge and authorization\footnote{Autonomy ensures independence from centralized authorities, while ownership and control grant users authority over credential use and sharing, with being being mutually dependent.}.    
        
        \subsubsection{Relevance to Digital Identity}

        These SSI principles form a single, unified property in our research due to their strong interconnection, independence, and relevance when examining challenges and eIDAS 2.0 compliance. Ultimately, this property indicates that the user is the sole entity governing and controlling their digital presence. 
Traditional password-based authentication  fail to satisfy these requirements. Although users seem to control their accounts by holding the password, providers still store identifying data, keep sensitive personal information, and can technically impersonate users because they control both identity and authentication. They are basically centralized identity infrastructures just like CA-based PKI systems. Because identity providers act as the single source and final authority over identity data, users can not meaningfully exercise ownership or control over how their information is processed.
Consent Management Platforms (CMPs) \cite{jha2025privacy} may sound compliant with at least control and consent principles, however, they do not since CMPs typically store consent records in centralized repositories and redistribute them to third-party data processors. Empirical studies show that many websites misuse these systems—placing cookies before consent, obscuring refusal options, or omitting consent mechanisms altogether—making user consent neither transparent nor enforceable \cite{nouwens2020dark}. Or recurring data breaches and surveillance-driven retention practices of major centralized providers. 

Both modern and traditional identity management have compliant systems.  Traditional graphical-signature–based identities exemplify this: individuals create and retain full control over  their own signatures as the single authoritative source for their use and disclosure. Traditional identity documents---such as identity cards---also satisfy this property. Their validity relies on the user’s physical possession of the document, which inherently expresses consent for its use. Although these documents are issued by official authorities, this requirement does not diminish their status as a single source of truth\footnote{Despite the existence of forgery threats, their integrity is maintained through secure materials, the introduction of photography, and the robustness of modern electronic signature schemes \cite{baechler2020document}.}; nonetheless, consent associated with such documents is usually limited to their physical presentation \cite{j2021role}.  Biometric-based identities can also comply with this property when implemented under strict constraints\footnote{Compliance requires that authenticators cannot access or retain raw biometric data beyond what is necessary for verification \cite{blanton2024privacy}.} \cite{blanton2024privacy, 10.1007/978-981-97-5081-8_23}.  

\subsubsection{Compatibility Assessment} 

     Several articles and directives explicitly aim to shift data and privacy control to users. According to Recitals (2-5\footnotemark{}  \footnotetext{Recital (2) from \href{http://data.europa.eu/eli/reg/2024/1183/oj}{Regulation (EU) 2024/1183}: "... the Commission to propose the development of a Union-wide framework ... to provide people with \textbf{control over their online identity and data} as well as to enable access to public, private and cross-border digital services."} \footnotemark{}  \footnotetext{Recital (3) from \href{http://data.europa.eu/eli/reg/2024/1183/oj}{Regulation (EU) 2024/1183}: "... to lead to wide deployment of a trusted, voluntary, \textbf{user-controlled digital identity} that is recognised throughout the Union and allows every user to control their data in online interactions." }
\footnotemark{} 
\footnotetext{Recital (5) from \href{http://data.europa.eu/eli/reg/2024/1183/oj}{Regulation (EU) 2024/1183}: "Union citizens ... should have the right to a digital identity that is \textbf{under their sole control} and that enables them to exercise their rights in the digital environment ..." } 
 \footnotemark{}
\footnotetext{Recital (4) from \href{http://data.europa.eu/eli/reg/2024/1183/oj}{Regulation (EU) 2024/1183}: "... The Declaration also states that everyone has the right to the protection of their personal data. \textbf{That right encompasses the control on how the data is used and with whom it is shared}." }), the EC is committed to ensuring citizens retain control over their eIDs. Recital (7)\footnotemark{}  \footnotetext{Recital (7) from \href{http://data.europa.eu/eli/reg/2024/1183/oj}{Regulation (EU) 2024/1183}: "... The European Digital Identity Framework is intended to achieve a shift from the reliance on national digital identity solutions only, to the provision of electronic attestations of attributes valid and legally recognized across the Union. ..." } indicates a goal to reduce exclusive reliance on government-issued eIDs by enabling the use of EU-wide electronic attestations of attributes. However, because electronic identification is legally tied to a PID, the mandated EUDIW requires eIDs under both the regulation and the ARF.

(Q)EEA providers also gain flexibility in setting rules for issuing and accepting attestations, while public administrations must accept any electronic format. The ARF envisions an EUDIW that gives users full control over their cryptographic keys and the use of PIDs and other attestations. However, this level of control remains limited: users cannot always access their wallet independently, as providers can revoke wallet instances. Such dependencies \textcolor{black}{partially impacts self-sovereign access} and security properties, especially since the wallet is meant to support secure local data storage \textcolor{black}{(A significant shift towards self-custody approach to manage credentials via WSCD/WSCA, is negatively impacted with a form of authoritarian wallet instance management)}. If a wallet instance is revoked, it may become unusable, preventing users from presenting attestations that do not even require a valid government eID. Moreover, although eIDAS 2.0 (Art. 5a\footnotemark{}\footnotetext{Letter a, point 4 of article 5a from : \href{http://data.europa.eu/eli/reg/2024/1183/oj}{Regulation (EU) 2024/1183}: "
European Digital Identity Wallets shall enable the user, in a manner that is user-friendly, transparent, and traceable by the user, to:
 (a). securely request, obtain, select, combine, store, delete, share and present, \textbf{under the sole control of the user}, person identification data and, where applicable, in combination with electronic attestations of attributes, to authenticate to relying parties ... "} \footnotemark{}\footnotetext{Point 14 of article 5.a, \href{http://data.europa.eu/eli/reg/2024/1183/oj}{Regulation (EU) 2024/1183}: "14.Users shall have \textbf{full control} of the use of and of the data in their European Digital Identity Wallet. "}) calls for user control to ensure that the EUDIW enables transparent exercise of data rights, this “control” aligns more with consent mechanisms than with genuine user authority over their digital identity. 

Although eIDAS 2.0 envisions the EUDIW as an independent source of identification data, it depends on activation through a valid PID. QTSPs issuing QEAAs must verify a user’s identity through legally recognized public-sector Authentic Sources (AS), creating a multi-step process in which user consent\footnotemark{}\footnotetext{Letter e, point 5 of article 5a from \href{http://data.europa.eu/eli/reg/2024/1183/oj}{Regulation (EU) 2024/1183}: " ... 5.European Digital Identity Wallets shall, in particular: (e)
in the case of the electronic attestation of attributes with \textbf{embedded disclosure policies}, implement the appropriate mechanism \textbf{to inform the user} that the relying party or the user of the European Digital Identity Wallet requesting that electronic attestation of attributes has the \textbf{permission to access such attestation};..."} is relayed through an AS intermediary\footnotemark{}\footnotetext{Recital (61) from \href{http://data.europa.eu/eli/reg/2024/1183/oj}{Regulation (EU) 2024/1183}: " ... 
Member States should establish appropriate mechanisms at national level to ensure that qualified trust service providers ... are able, on the \textbf{basis of the consent of the person to whom the attestation is issued}, to verify the authenticity of the attributes relying on authentic sources ..."} before attribute verification can occur. RPs request access to PID or (Q)EAAs attributes via the OIC for Verifiable Presentations (OID4VP) protocol, allowing users to consent to sharing specific information. However, the wallet can only be considered a true single source when credential verification does not require reference to an external authentic source, as dependence on intermediaries diminishes the user’s authority over their own identity data. \textcolor{black}{Although the underlying technology required to support this property, the trust model associate with issuance  and verification workflows structurally constrains ownership, control and the perceived legitimacy of credentials within the wallet. Nonetheless, non-qualified credentials can be fully compliant, provided  that their verification methods are  not restricted EUDI wallet providers}.

\subsubsection{Challenges \& gaps}
\begin{itemize}
     \item The Exclusive reliance on government issued PIDs for authentication \textcolor{black}{may potentially} restrict access to EUDIW services.  
     \item The dependence on specific EUDIWPs when using identity services \textcolor{black}{undermines} user control \textcolor{black}{ and ownership, especially if their authority is abused}. 
     \item The reliance on external authorities providing authentic sources renders the EUDIW not a direct single source of authentic identity data. 
\end{itemize}

 \subsubsection{Recommended improvements}

\begin{itemize}
        \item  \textcolor{black}{Recommend}  and adopt DID ownership proofs as the basis for identity on-boarding, binding eIDAS PIDs to user-controlled DIDs\footnote{PID issuance should require verifiable proof of DID control, either through in-person verification or privacy-preserving remote mechanisms such as ZK-biometric authentication \cite{10.1007/978-981-97-5081-8_23} or homomorphic encryption \cite{9363426}.}.       
 
    \item Ensure the ability to use different EUDIWs relying on DID-anchored credentials, by supporting secure import/export functionalities secured with user's DID ownership verification. 
    
 \item Ensure (Q)EAAs issuance is consented and verifiable using DID signatures of both issuer and identity holder. As well as the uage of SD-VP in order to control data disclosure  and strengthen accountability with non-repudiation properties.
 \item Use public, EU-governed infrastructure for trust anchoring and credential revocation using optimized hybrid transactions. Such network could be a layer-2 solution (e.g., via rollups \cite{katsika2024critical} or validium \cite{lavin2024survey})\footnote{A \textit{rollup} is a layer-2 scaling solution that processes transactions off-chain and periodically submits proofs to the main blockchain. Rollups can be \textit{optimistic}, assuming transactions are valid unless challenged, or \textit{zero-knowledge} (zk-rollups), using cryptographic proofs (e.g., SNARKs) for verification. A \textit{validium} is similar but stores transaction data off-chain, improving scalability while relying on cryptographic commitments for security, which could allow EU member states to erase historic transaction data if required.} that ensures credentials verification is completed directly from the EUDIW as a single source of truth.  
  \end{itemize}

\begin{figure}[h]

\centering 
    \includegraphics[width=0.95\textwidth]{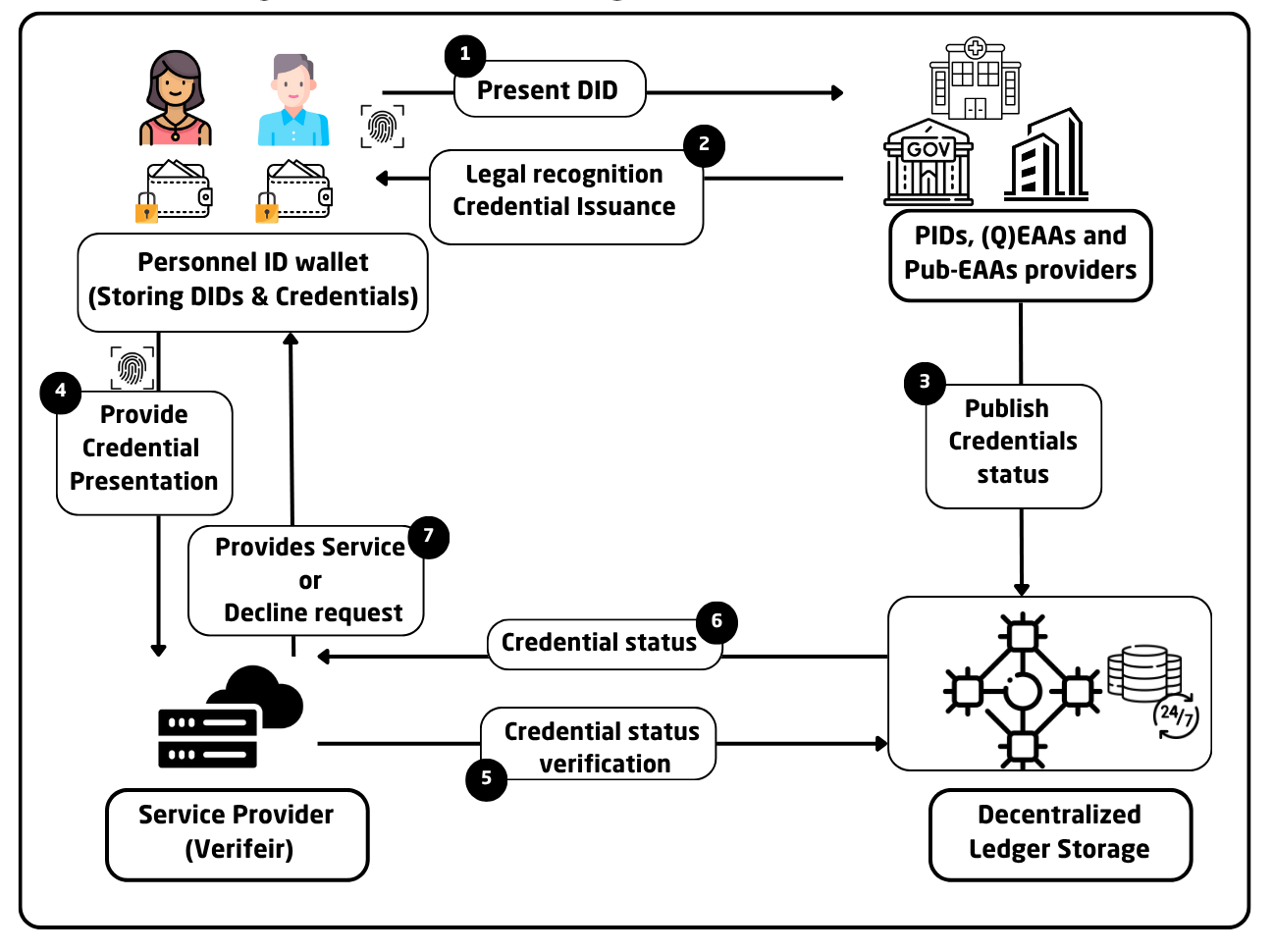}
    \caption{ DID-VC system interaction model showing how users can preserve their autonomy, control, ownership and single source properties.} 
 
    \label{did/vc 1}
    \end{figure}

 A DID architecture relies on a public network that eliminates centralized intermediaries in credential usage, avoiding redirection to issuers for verification.  Figure \ref{did/vc 1} illustrates how users can authenticate their credentials directly with service providers, eliminating centralized intermediaries and ensuring greater autonomy and control over their digital interactions

 \subsection{Privacy and Minimal Disclosure}

        \subsubsection{Definition }
        The individual’s ability to sovereignly manage the disclosure of personal identity data by revealing only the minimum information necessary for a given interaction. Users must be able to authenticate  using selectively disclosed or anonymized attributes when full legal identification is not required. The identity ecosystem must technically and legally enforce data minimization, ensuring that mandatory identification processes---particularly in government and regulated services---do not exceed their legitimate purpose.

      \subsubsection{Relevance to Digital Identity}

     Privacy and minimal disclosure are satisfied by only a limited set of existing systems, often under restrictive assumptions. Most traditional identity systems fail to meet these properties: physical identity documents (e.g., national IDs or driver’s licenses) require full disclosure even when only a single attribute is needed, while digital certificate systems such as X.509 \cite{berbecaru2023evaluation} and PGP (Pretty Good Privacy) \cite{mathew2021can} cryptographically bind identity attributes as indivisible units, preventing selective disclosure without invalidating the signature. Complying solutions partially address this limitation through privacy-preserving cryptographic techniques, such as HE-based biometric authentication \cite{alberto2015privacy, 9363426,  yang2023review}, which enable authentication or verification without revealing raw personal data. Additional approaches based on pseudonymisation \cite{stalla2016anonymous} and anonymization offer limited privacy guarantees but suffer from linkability or loss of reusability and accountability, constraining their applicability in high-assurance or regulated settings. 

     Techniques such as hash-based, signature-based, and zero-knowledge methods \cite{RAMIC2024916}, are primarily designed for SSI ecosystems and are   uncommon in centralized systems, while simpler approaches like  \cite{chadwick2019verifiable} are inefficient to deploy at full scale due to their operational and management overhead.

     \subsubsection{Compatibility Assessment} 

      eIDAS 2.0 strongly emphasizes selective disclosure \footnotemark{}\footnotetext{Recital (15) from \href{http://data.europa.eu/eli/reg/2024/1183/oj}{Regulation (EU) 2024/1183}: "... All Union citizens ... should be empowered to securely request, select, combine, store, delete, share and present data related to their identity and request the erasure of their personal data in a user-friendly and convenient way, under the \textbf{sole control of the user}, while enabling {selective disclosure} of personal data. ... "}
\footnotemark{}\footnotetext{Recital ( 59) from \href{http://data.europa.eu/eli/reg/2024/1183/oj}{Regulation (EU) 2024/1183}: "...The European Digital Identity Wallet should technically enable \textbf{the selective disclosure of attributes} to relying parties. It should be technically possible for the user \textbf{to selectively disclose attributes}, including from multiple, distinct electronic attestations, and to combine and present them seamlessly to relying parties. This feature should become a basic design feature of European Digital Identity Wallets, ..."} but treats it as an optional service rather than a mandatory user right, meaning that eID schemes and  EUDIWs can be certified even without supporting selective disclosure.  This is reflected   in article 5.a 
\footnotemark{}\footnotetext{Point 5, letter (a) of article 5a from \href{http://data.europa.eu/eli/reg/2024/1183/oj}{Regulation (EU) 2024/1183}: " 5. The European Digital Identity Wallet shall , a) support common protocols and interfaces : ... (iii)
for the sharing and presentation to relying parties of person identification data, electronic attestation of attributes or of \textbf{selectively disclosed} related data online and, where appropriate, in offline mode; ... }  defining selective disclosure as a supported service within the EUDIW common protocols.  The ARF proposes attestation formats that support data minimization, notably Selective Disclosure for JWTs \cite{wu2025selective} (\textit{SD-JWT}), which enables secure and verifiable selective disclosure of credential attributes. SD-JWT–based Verifiable Credentials are compatible with W3C standards \cite{mazzocca2025survey}.
\\ EUDIW users can apply these formats when presenting PIDs or attestations to RPs, provided that issuers embed appropriate disclosure policies. Effective user-controlled disclosure therefore depends on both issuers and RPs enabling minimization, highlighting the need for clear regulation defining when and on what grounds RPs may request personal data.

Untraceability is a key privacy principle, protecting users from tracking and documentation of their activities \cite{garcia2021towards, bussard2004untraceable}. While traceability is often used for advertising, accountability, or security---and sometimes to intimidate users---sensitive information like attestation status, eID or EUDIW validity, and other identity data should remain untraceable to preserve privacy. However, eIDAS 2.0 and the ARF currently do not address this, and the challenge is complex, as traceability is sometimes necessary for security or liability purposes.

\subsubsection{Challenges \& gaps}
\begin{itemize}
    \item The optional treatment of selective disclosure in eIDAS 2.0 weakens user sovereignty, as certification does not require this capability.
    \item The reliance on PID/attestation providers and RPs to enable disclosure policies limits user autonomy and weakens effective data minimization.
    \item The lack of untraceability safeguards for attestations, PIDs, and EUDIWs compromises the privacy of digital identity.
\end{itemize}

 \subsubsection{Recommended improvements}
 Although the regulation does not explicitly conflict with this property, clarifying and strengthening it would better safeguard privacy: 
 \begin{itemize}
\item Require that EUDIWs and compatible identity systems support selective disclosure–capable credential formats (e.g., SD-JWT VCs) by design.

\item Enable users to autonomously generate multiple PID-bond DIDs---only when official identification is explicitly required, allowing the separation of contexts and preventing cross-service identity correlation.

\item Allow issuers to designate which credential claims are eligible for selective disclosure, while ensuring that users retain full control over which eligible claims are disclosed to verifiers.

\item Ensure that selectively disclosed claims are protected against inference and dictionary attacks through cryptographic salting and hashing mechanisms.
 
\item Ensure that untraceable mechanisms are enforced to safeguard credential's usage history and status \cite{sitouah2024untraceable, guo2021uppresso, alupotha2025anonymous}. While providing identity with usage history by RPs \cite{yeoh2023fast, rabzelj2025beyond}.  
\end{itemize}

\subsection{Verifiability and Authenticity}
            \subsubsection{Definition }
            
This refers to the ability of users to reliably prove that their digital identities and associated attributes are genuine, accurate, and under their legitimate control. The identity system must provide cryptographically verifiable proofs that ensure the authenticity and integrity of identification data. Where anonymisation or pseudonymisation is applied, the system must still guarantee that the authenticated identity corresponds to a real natural or legal person.
      \subsubsection{Relevance to Digital Identity}
      
Verifiability and Authenticity are essential for every IDMS relying on cryptographic verification and controlled authentication processes, supported for both on-site and remote verification by either physical inspection, biometric matching, or digital signature validation against the identity's issuer database and revocation lists. By contrast, traditional paper-based identity documents and handwritten signatures fail to provide strong verifiability guarantees. Physical documents depend on assumptions about material scarcity and the difficulty of forgery. 
      
     \subsubsection{Compatibility Assessment}

   Decentralized Identity Models and the proposed EU eIDs\footnotemark[46] 
 \footnotemark[43] 
\footnotemark{}\footnotetext{Number 2 of article 5f from: \href{http://data.europa.eu/eli/reg/2024/1183/oj}{Regulation (EU) 2024/1183}: "Where private relying parties that provide services ... are required by Union or national law to use strong user authentication for online identification or where strong user authentication for online identification is required by contractual obligation, ...  "} share similar principles for authenticity, where the latter use a Presentation Interface through which RPs request and receive PIDs or (Q)EAAs via the corresponding presentation protocol---OID4VP for EUDIWs, typically implemented as a web or mobile app compliant with ISO/IEC 18013-5. Attestation presentation between an RP and a user’s wallet follows four defined modes\footnote{(1) Proximity Supervised Flow (PSF): in-person exchange using short-range technologies (e.g., NFC, Bluetooth) under human supervision; (2) Proximity Unsupervised Flow (PUF): similar proximity-based exchange without supervision; (3) Remote Cross-Device Flow (RCDF): the user accesses service information on a separate device (e.g., via QR code) while the wallet authenticates the session; (4) Remote Same-Device Flow (RSDF): the wallet and service interaction occur on the same device, securing the session directly.} enabling both proximity-based and remote interactions.
Each Member State must operate a publicly available compliant RP registrar\footnotemark{}\footnotetext{Recital (18) from \href{http://data.europa.eu/eli/reg/2024/1183/oj}{Regulation (EU) 2024/1183}: "... supervisory bodies ... should, upon notification, ... the inclusion of relying parties in the authentication mechanism are withdrawn or suspended until the notifying authority confirms that the irregularities identified have been remedied. "} \footnotemark{}\footnotetext{Number 5 of article 5b form:\href{http://data.europa.eu/eli/reg/2024/1183/oj}{Regulation (EU) 2024/1183}: "
5.Member States shall make the information referred to in paragraph 2 publicly available online in electronically signed or sealed form suitable for automated processing...\\
9.Relying parties shall be responsible for carrying out the procedure for authenticating ... Relying parties shall not refuse the use of pseudonyms, where the identification of the user is not required by Union or national law.
"} secured by a Certificate Authority, with the EUDIW verifying certificates of RPs and PID/(Q)EAA providers to establish legitimacy and enable a trusted authentication framework.

The ARF deliberately omits specifications for verifying users with inactive EUDIWs\footnote{Authentication and submission of a valid PID activate the wallet.}, leaving implementation to PID providers and Member States due to variations in national identification systems. Verification of attestations relies on provider-generated signatures or seals, or alternatively on device-based or self-signing mechanisms when the EUDIW confirms the user holds a WSCD/WSCA at a high LoA. Ultimately, acceptance of these certificates is determined by the RP’s verification policies\footnotemark{}\footnotetext{Letter c of article 1 from 
\href{http://data.europa.eu/eli/reg/2024/1183/oj}{Regulation (EU) 2024/1183}: " ..., this Regulation:
(c)
establishes a legal framework for electronic signatures, electronic seals, electronic time stamps, electronic documents, electronic registered delivery services, certificate services for website authentication, electronic archiving, electronic attestation of attributes, electronic signature creation devices, electronic seal creation devices, and electronic ledgers.’; "}. 

eIDAS 2.0 \textcolor{black}{promotes} the use of pseudonyms\footnotemark{}\footnotetext{Recital (22) from \href{http://data.europa.eu/eli/reg/2024/1183/oj}{Regulation (EU) 2024/1183}: "... Wallets should include a functionality to generate user-chosen and managed pseudonyms, to authenticate when accessing online services. "} \footnotemark[55] 
to support user anonymity during interactions with RPs in accordance with the GDPR. Furthermore, the ARF designs an EUDIW with an  operational status using  backend-based user authentication mechanisms (e.g., email) without a verified PID, resulting in a   wallet with limited functionality. However, if wallet providers are able to observe and correlate usage of such pseudonymous credentials at the wallet-unit level, these credentials may become effectively linkable over time. As a result, only fully operational wallets that are designed to prevent such correlation can provide meaningful anonymity guarantees. In practice, achieving strong anonymity alongside simultaneously supporting fully identified credentials within the same wallet environment remains technically and architecturally constrained.
\subsubsection{Challenges \& gaps}
\begin{itemize}
    \item The reliance on a certificate-based RP registrar and CA-managed trust framework reinforces centralized control, restricting EUDIW interactions to pre-approved service providers. 
    \item The requirements imposed on pseudonymisation and anonymisation make these options traceable by either the EUDIW or attestation providers.
\end{itemize}
 \subsubsection{Recommended improvements}
\begin{itemize}
\item Ensure that attestation are cryptographically bound to DIDs, enabling independent identity verification using the DID's private key.
\item \textcolor{black}{Recommend}  users to verify the authenticity and integrity of credentials autonomously, without requiring real-time access to issuers or RPs.
\item Require that attestation  presentation requests be signed and cryptographically attributable to their respective issuers, holders, and verifiers.
\item Support public, tamper-evident registries for publishing credential status information (e.g., issuance, transfer, revocation) to enable third-party verification.

\item \textcolor{black}{Recommend}  credentials to operate as transferable or updatable digital assets when appropriate, while preserving authenticity via cryptographic witness proofs \cite{chakraborty2023map}.
\item Reduce dependency on centralized authorities by ensuring that authenticity derives from cryptographic proofs and public proof rather than institutional trust alone.
\end{itemize}

 \subsection{Security and Protection}

            \subsubsection{Definition } This refers to the ability of the digital identity ecosystem to safeguard identities, credentials, and interactions against unauthorized access, misuse, and malicious activities through the application of state-of-the-art cryptographic and security mechanisms\footnote{Complying to international security and privacy standards \cite{taherdoost2022understanding} such as ISO27001, NIST, SOC2, GDPR ... etc}. Identity systems must ensure strong authentication and authorization of users, secure and resilient communication channels for identity data exchange, and continuous protection against threats through preventive and detective controls. Solution providers are required to comply with recognized security standards \cite{taherdoost2022understanding}, undergo regular certification and audits, and maintain mechanisms that ensure accountability, non-repudiation, and protection of users from fraudulent, coerced, or unauthorized actions.

      \subsubsection{Relevance to Digital Identity}

      Digital identity resilience is paramount with information technology across the board. As up-to-date cryptographic mechanisms and secure operational practices are a must. Many contemporary identity systems satisfy deploy security frameworks such as  standardized Authentication, Authorization, and Accounting (AAA) protocols \cite{metz2002aaa} which enforce strong authentication, granular authorization, and auditable accounting of user actions. When implemented in compliance with international standards (e.g., ISO/IEC 27001), these systems provide protection against unauthorized access and enable evidence of interactions. 

Conversely, several widely deployed authentication mechanisms fail to meet security and protection requirements. Basic authentication schemes, such as HTTP Basic Authentication \cite{naylor2014cost} or weak Extensible Authentication Protocol (EAP) variants (e.g., EAP-MD5, EAP-TTLS/PAP) \cite{dantu2007eap, malekzadeh2009vulnerability}, rely on outdated cryptographic assumptions, lack mutual authentication, or expose credentials to replay and man-in-the-middle attacks.  
     \subsubsection{Compatibility Assessment} 

     eIDAS 2.0 advances the security objectives traditionally associated with an IDMS by mandating high assurance levels, strong cryptographic controls, and regulated interoperability within the EUDIW ecosystem. The ARF emphasizes secure interoperability through certified cybersecurity schemes and conformity assessments performed by accredited CABs. Although no dedicated wallet security schemes exist yet under the EU Cybersecurity Act (CSA), the regulation requires technologies to achieve the highest levels of security, privacy  and enforces assurance level high (LoA high) for identity onboarding\footnotemark{}\footnotetext{Recital (15) from \href{http://data.europa.eu/eli/reg/2024/1183/oj}{Regulation (EU) 2024/1183}: "... Technologies used to achieve those objectives should be developed aiming towards the \textbf{highest level of security, privacy}, ... "}\footnotemark{}\footnotetext{Recital (28) from \href{http://data.europa.eu/eli/reg/2024/1183/oj}{Regulation (EU) 2024/1183}: "... The onboarding of Union citizens ... should be facilitated by relying on electronic identification means issued at assurance level high. Electronic identification means issued at assurance level substantial should be relied upon only where harmonised technical specifications and procedures using electronic identification means issued at assurance level substantial in combination with supplementary means of identity verification will allow the fulfillment of the requirements set out in this Regulation as regards assurance level high. ... To ensure sufficient uptake of European Digital Identity Wallets, harmonised technical specifications ... including those issued at assurance level substantial, should be set out in implementing acts.. "}, supplemented by additional verification mechanisms when lower assurance levels are used.

Security is further reinforced through architectural elements such as cryptographic key–PID binding and validation, WSCDs/WSCAs, QES services provided by QTSPs under  European Telecommunications Standards Institute (ETSI) standards, and a managed wallet instance lifecycle enabling revocation in case of compromise. Protection of users is primarily addressed through centralized accountability: EUDIW providers and QTSPs must undergo periodic conformity assessments, are subject to supervisory oversight, and are legally liable for damages arising from non-compliance\footnotemark{}\footnotetext{Number 4 of article 46a  from \href{http://data.europa.eu/eli/reg/2024/1183/oj}{Regulation (EU) 2024/1183}:  
4.Qualified trust service providers that have been granted their qualified status under this Regulation before 20 May 2024 shall submit a conformity assessment report to the supervisory body proving compliance with Article 24(1), (1a) and (1b) as soon as possible and in any event by 21 May 2026.’; "} \footnotemark{}\footnotetext{Number 1 or article 21 from \href{http://data.europa.eu/eli/reg/2024/1183/oj}{Regulation (EU) 2024/1183}: "‘1.Where trust service providers intend to start providing a qualified trust service, they shall notify the supervisory body of their intention together with a conformity assessment report issued by a conformity assessment body confirming the fulfillment of the requirements laid down in ... "}\footnotemark{}\footnotetext{Number 1 of article  13 from \href{http://data.europa.eu/eli/reg/2024/1183/oj}{Regulation (EU) 2024/1183}: "...trust services shall be liable for damage caused ... due to failure to comply with the obligations under this regulation. in Any natural or legal person who has suffered material or non-material damage as a result of an infringement of this Regulation by a trust service provider shall have the right to seek compensation in accordance with Union and national law. "}.

However, while eIDAS 2.0 ensures institutional trust, legal enforceability, and high technical security, it only partially conforms to the SSI protection property. Users do not possess independent, cryptographic means to prove violations of access, data misuse, or denial of service without relying on provider-controlled logs and authorities. The absence of transparency mechanisms such as public ledgers limits user-verifiable accountability, meaning that protection is achieved through regulation and liability rather than through full user-centric, self-verifiable control.

\subsubsection{Challenges \& gaps}
\begin{itemize}
    \item The lack of independent, cryptographic means to prove violations (access, data misuse, or DoS), relying instead on provider-controlled logs.
    \item The absence of transparency mechanisms limits user-verifiable accountability and visibility of potential breaches.
    \item The over-reliance on regulated entities for compliance and accountability, which reduces user control over security and protection.
    \item The lack of dedicated wallet security schemes under the EU (CSA) leaves gaps in achieving consistent, high security standards for digital wallets.
\end{itemize}
 \subsubsection{Recommended improvements} 
To ensure robust security and effective protection of users’ identities and rights:

\begin{itemize}
\item Require the use of strong, up-to-date cryptographic algorithms for key generation, signing, encryption, and verification.
\item Enable users to generate and manage cryptographic keys locally and autonomously, ensuring  control over identifiers and credentials.
\item Mandate that issuers and verifiers accept only credential formats and verification methods that rely on publicly auditable and secure cryptographic standards.
\item \textcolor{black}{Recommend}    transparent and verifiable security mechanisms by favoring open-source implementations of DID, VC, and selective disclosure schemes.
\item Allow users to detect, log, and audit access to their identity data without relying on intermediaries, enabling independent evidence of interactions.
\item Leverage decentralized ledgers to protect data integrity and availability while minimizing single points of failure.
\item \textcolor{black}{Recommend}  cryptographic protection of user rights through mechanisms such as zero-knowledge proofs and smart contracts, ensuring fairness and non-discrimination without exposing sensitive data.
\end{itemize}

 \subsection{Accessibility and Availability }
            \subsubsection{Definition }
            Accessibility and Availability refer to the individual’s ability to consistently and securely access their own identity information and attestations at any time and across platforms, without reliance on or interference from intermediaries. This includes access to verifiable records of updates or changes, while ensuring exclusive access to one’s own data in line with autonomy and control.
            
      \subsubsection{Relevance to Digital Identity}

Accessibility and Availability are properties that require users to retain continuous access to their identity data and credentials, independent of service outages or intermediary control. Biometric passports \cite{chung2024will} illustrate a system that largely satisfies this property: once issued, the holder maintains direct access to their identity data embedded in the eMRTD (electronic Machine-Readable Travel Document), which can be authenticated offline using digitally signed data. While revocation may restrict certain privileges, it does not eliminate access to the underlying identity information or the ability to be authenticated.

In contrast, centralized identity management systems fail to fully satisfy accessibility and availability requirements. In these systems, identity data is stored and controlled by service providers, and user access is mediated exclusively through provider-managed interfaces. Availability is therefore contingent on the provider’s infrastructures. As a result, users cannot reliably access their identity data independently of the provider.

     \subsubsection{Compatibility Assessment} 

eIDAS 2.0 defines accessibility as a right for natural and legal persons, including the hardware and software needed for identity services\footnotemark{}\footnotetext{article 12b from \href{http://data.europa.eu/eli/reg/2024/1183/oj}{Regulation (EU) 2024/1183}: "... Regulation, gatekeepers shall in particular allow them effective interoperability with, and, for the purposes of interoperability, \textbf{access} to, the same operating system, hardware or software features. Such effective interoperability and \textbf{access shall be allowed free of charge} and regardless of whether the hardware or software features are part of the operating system, \textbf{are available} to, or are used by, that gatekeeper when providing such services.... "} \footnotemark[49]. 
 However, unlike SSI, the regulation does not discourage intermediaries. Instead, the ARF proposes verifying certificates (for providers and RPs), attestations (including PIDs), and wallet instances through issuer-provided platforms. This means that when Emma uses her EUDIW to verify an RP’s legitimacy, she must rely on the relevant member state’s RP register and certificate authority. Similarly, a municipal office must verify the validity of Emma’s wallet with her EUDIW provider and then check her PID or other attestations. As a result, all participants depend on trusted intermediaries and their interfaces to complete these procedures.

 Accessibility would be less problematic if eIDAS 2.0 adopted a decentralized approach to ensure high availability; however, its current version does not actively encourage the use of decentralized technologies such as DLTs or blockchains. Moreover, the ARF does not address the scalability of EUDI schemes, despite earlier eIDAS efforts falling short at the EU level due to security and scalability limitations. Although EUDIW aims for high conformity and interoperability across vendors and Member States, prioritizing scalability is essential for a truly large-scale, interoperable, and portable deployment.

\subsubsection{Challenges \& gaps}
\begin{itemize}
    \item The reliance on issuer-provided platforms for certificate, attestation, and wallet verification introduces intermediary dependencies.
    \item The ARF does not address the scalability requirements of EUDI schemes.
      \item The lack of emphasis on decentralized technologies (e.g., DLTs) reduces resilience and high-availability guarantees.

\end{itemize}
 \subsubsection{Recommended improvements}
 \begin{itemize}
    \item Avoid reliance on centralized service providers by supporting decentralized verification of credentials, registries, and wallet instances.
    \item Enable users to access, store, and verify DIDs, VCs, and revocation data locally without dependency on issuer---or state---controlled platforms.
  \item \textcolor{black}{Recommend}  peer-to-peer communication models between users, issuers, and RPs to ensure uninterrupted access during partial network or service outages, downtime, and  denial-of-service attacks.
    \item Ensure that no single authority can unilaterally deny users access to identity verification or credential validation services.

 \end{itemize}

 \subsection{Recoverability, Persistence and Interoperability}
            \subsubsection{Definition } 
            The combination of these properties refers to the ability of digital identities and their associated credentials to be reliably recovered, maintained over time, and used across platforms\footnote{The identity ecosystem enables portability and cross-domain interoperability through open, license-free standards, ensuring that identities and credentials can be securely represented, exchanged, and verified without binding users to a specific provider or technological silo.}, domains, and service providers without loss of control by the identity holder. Identity systems must enable identity holders to autonomously recover or recreate self-generated credentials and cryptographic materials through user-defined recovery mechanisms. Identities should be persistent and long-lived where appropriate, while permitting controlled modification, revocation, or invalidation under predefined conditions that preserve user autonomy.

      \subsubsection{Relevance to Digital Identity}
      
      Recoverability, Persistence, and Interoperability constitute a core SSI property that is largely unmet---simultaneously---by conventional identity systems, which tightly couple identity lifecycle, continuity, and usability to specific infrastructures or providers. Traditional cryptographic identity models, such as self-signed certificate authorities \cite{berkowsky2017security}, fail to ensure recoverability: the loss of a root private key  irreversibly compromises the entire identity domain, rendering all dependent identities unusable. Similarly, social media and email-based identities lack persistence, as their existence is contingent on the continued operation and policies of service providers \cite{greschbach2012devil}. Similarly, national eID schemes and commercial proprietary identity systems rely on heterogeneous technical and procedural frameworks that prevent seamless cross-domain use, limiting  portability and user autonomy. These limitations demonstrate how these identities'  lifecycle control are subordinated to issuers and service providers rather than to identity holders.

      Within centralized or provider-controlled paradigms, The following examples illustrate systems that satisfy one or more of the properties of recoverability, persistence, and interoperability, though none satisfy all three simultaneously. Mnemonic seed phrases \cite{woo2016improving} often used in non-custodial cryptocurrency wallets, offer a stronger form of user-controlled recoverability and portability. When securely stored, seed phrases allow users to restore identities and assets\footnote{This approach introduces an irreversible failure mode: the loss of the recovery phrase results in permanent loss of identity. Custodial wallets mitigate this risk by offering account recovery services, but at the cost of transferring control and trust to the custodian \cite{seymour2024custodial}.}. This cryptographic mechanism can still be adopted by centralized systems such as PKI forming a hybrid system with better recovery and persistence \cite{singh2021private, chatzigiannis2023sok}. Although many existing identifiers are long-lived by design (e.g., university diplomas, birth certificates, DOIs, ORCIDs), they remain valid only as long as a trusted authority or foundation maintains them, lacking an independent technology that transfers persistence to the identity holders themselves.

     \subsubsection{Compatibility Assessment} 

     EUDIWs rely on notified eID schemes, for which recoverability after unexpected events is a legal obligation of both providers and authorities. However, this aspect is currently outside the scope of the ARF or has not yet been explicitly addressed. The only related provisions regulated under eIDAS 2.0 concern the use of archiving solutions to ensure data preservation\footnotemark{}\footnotetext{Letter c, Number 1 of article 45j from \href{http://data.europa.eu/eli/reg/2024/1183/oj}{Regulation (EU) 2024/1183}: "1. Qualified electronic archive services shall meet the following requirements: ... (c)
they ensure that those electronic data and those electronic documents are preserved in such a way that they are safeguarded against loss and alteration... "}and the immediate revocation of an EUDIW in cases of loss or theft\footnotemark{}\footnotetext{Recital (34) from \href{http://data.europa.eu/eli/reg/2024/1183/oj}{Regulation (EU) 2024/1183}: "... Member States should develop simple and secure procedures for the users to request immediate revocation of validity of European Digital Identity Wallets, including in the case of loss or theft ... "}.  

While certain interactions legitimately require higher assurance levels, the regulation’s allocation of full liability to PID and EUDIW providers conflicts with the concept of digital identity as an extension of EU citizenship rights\footnote{Art 18, The Treaty on the Functioning of the European Union and Ch V, the Charter of Fundamental Rights on   EU citizenship rights.}\footnote{ch 4.6.3 Wallet Unit/Instance, v2.8.0 of eIDAS 2.0 Architecture and Reference framework}. \textcolor{black}{Other non-qualified credentials can fully support persistence as long as their issuance method fully support recoverability, portability and decentralized verification.} 

As repeatedly stated in this article, the regulation strongly emphasizes interoperability objectives\footnotemark{}
 \footnotetext{Recital (8) from \href{http://data.europa.eu/eli/reg/2024/1183/oj}{Regulation (EU) 2024/1183}: "... Additionally, investments have been made in both national and cross-border solutions ... including the \textbf{interoperability of notified electronic identification schemes} ... "}\footnotemark{}\footnotetext{Recital (19) from \href{http://data.europa.eu/eli/reg/2024/1183/oj}{Regulation (EU) 2024/1183}: "... and technical specifications to ensure \textbf{seamless interoperability and to adequately increase IT security}... "}.
 The vision\footnotemark{}\footnotetext{Recital (11) from \href{http://data.europa.eu/eli/reg/2024/1183/oj}{Regulation (EU) 2024/1183}: "... European Digital Identity Wallets should facilitate the application of the ‘once only’ principle, thus reducing the administrative burden on and supporting \textbf{cross-border mobility} ... and businesses across the Union and fostering the development of \textbf{interoperable e-government services across the Union}. "} 
   \footnotemark{}\footnotetext{Recital (42) from \href{http://data.europa.eu/eli/reg/2024/1183/oj}{Regulation (EU) 2024/1183}: "... For the convenience of users and in order to ensure \textbf{cross-border availability of such services}, it is important to undertake actions  ... Such guidelines should be prepared taking into account the interoperability framework of the Union. Member States should have a leading role when it comes to adopting those guidelines.. "}  \footnotemark[60] 
 for EU cross-border seamlessness is a federated ecosystem in which multiple frameworks operate like a single sign-on system, allowing all notified eID schemes to function as both identity providers and SPs. While the recitals\footnotemark{}\footnotetext{Recital (36) from \href{http://data.europa.eu/eli/reg/2024/1183/oj}{Regulation (EU) 2024/1183}: "... To ensure that the European Digital Identity Framework is open to innovation, ... Member States are encouraged, jointly, to set up sandboxes to test innovative solutions in a controlled and secure environment in particular to improve the functionality, \textbf{protection of personal data, security and interoperability of the solutions} and to inform future updates of technical references and legal requirements. That environment should foster the inclusion of SMEs, start-ups and individual innovators and researchers, as well as relevant industry stakeholders. ... "} propose controlled environments for joint testing with industry and researchers, similar proposals under the original regulation were never realized due to the absence of binding provisions, although multiple implementing regulations\footnote{COMMISSION IMPLEMENTING REGULATION (EU) C(2024) 8495: rules for the integrity and core functionalities of eID Wallets, C(2024) 8496: rules on the protocols and interfaces of eID Wallets solutions, C(2024) 8498: rules on person identification data and electronic attestations of attributes of eID Wallets, C(2024) 8507: reference standards, specifications and procedures for a certification framework for eID Wallets, C(2024) 8516: obligations for notifications to the Commission concerning the eID Wallet ecosystem } restricts providers to particular format (e.g., PID and QEEA must comply to ISO/IEC.18013-5:2021 or Verifiable Credentials Data Model 1.1 \footnote{COMMISSION IMPLEMENTING REGULATION (EU) C(2024)2979 - 28 November 2024 }).

\subsubsection{Challenges \& gaps}
\begin{itemize}
    \item The regulation lacks clear provisions for credential recovery and user remediation in failure or compromise scenarios.
    \item The continuous validation requirement imposed on PID and EUDIW providers undermines user control and identity persistence.

\end{itemize}
 \subsubsection{Recommended improvements}
\begin{itemize}
    \item \textcolor{black}{Recommend}  non-custodial key generation and \textcolor{black}{credential} management\footnote{This allows  users to create, back up, and recover DIDs using mnemonic phrases \cite{woo2016improving}, Hardware Security Modules (HSMs), or equivalent secure mechanisms \cite{cirne2024hardware}.}.
    \item Mandate credential backup and recovery mechanisms that do not rely on centralized authorities such as encrypted replications, social recovery \cite{pedin2023smart, kat2025anarkey}, hardware-assisted \cite{malamas2025ha, cirne2024hardware}, or multi-wallet recovery strategies \cite{yu2024don}.

    \item Decouple identifier persistence from key control by supporting long-lived DIDs that remain resolvable even if private keys are lost.
    \item \textcolor{black}{Recommend} long-lived heterogeneous credentials and public witnesses on immutable networks to ensure verifiability independent of issuer availability\footnote{In the context of VCs and DLTs, a witness is cryptographic evidence that proves a credential’s existence, authenticity, and integrity at a specific point in time \cite{chakraborty2023map}. It is typically a cryptographic commitment (e.g., a hash or accumulator) that allows verification without contacting the original issuer. The witness acts as an immutable reference, enabling credential verification even if the issuer is unavailable or no longer operational.}.

    \item Promote open, publicly specified DID and VC standards to ensure cross-platform and cross-border interoperability, while avoiding binding identity usage to a single wallet, protocol, or vendor by supporting multiple EUDIW implementations and VC formats.
 
\end{itemize}

 \section{Discussion and Practical Implications}
 \label{Findings and Discussion}

In this paper, we show that eIDAS 2.0 regulation seeks to harmonize electronic identification and digital transactions across the EU to enable effective cross-border interoperability, while the ARF is intended to serve as a technical reference that minimizes divergence and interpretative ambiguity. Nevertheless, a number of core functionalities remain subject to implementation choices by individual Member States. This design choice represents a double-edged sword: on the one hand, greater technical inclusiveness may enhance interoperability \cite{vano2025social, loutocky2025certification, bukhari2024defining}; on the other hand, stricter standardization through the enforcement of specific norms and protocols may lead to stronger harmonization \cite{podda2025impact, babel2025self, lepore2024aligning, gosslbauer2025towards}. More specifically, the desirable outcome is standardization that remains highly accessible and open source, supporting both large and small organizations while remaining affordable for end users.
Building on this tension, our analysis in Section \ref{Analysis: SSI Property} identifies gaps and constraints in order to provide a balanced evaluation and formulate recommendations aligned with SSI principles in the context of eIDAS 2.0 and the EUDIW, \textcolor{black}{summarized in Table. \ref{tab:ssi-improvements}: 
} 

{ \color{black}
\setlength{\LTleft}{0pt}
\setlength{\LTright}{0pt}
\setlength{\tabcolsep}{4pt}
\renewcommand{\arraystretch}{1.05}
\footnotesize
\begin{longtable}{@{}>{\raggedright\arraybackslash}p{3.00cm}>{\raggedright\arraybackslash}p{5cm}>{\raggedright\arraybackslash}p{5cm}@{}}
\caption{Summary of recommended improvements for SSI alignment in the EUDI framework.}
\label{tab:ssi-improvements} \\
\hline
\textbf{SSI properties} & \textbf{Technical recommendation} & \textbf{Legal recommendations} \\
\hline
\endfirsthead

\hline
\textbf{SSI properties} & \textbf{Technical recommendation} & \textbf{Legal recommendations} \\
\hline
\endhead

\hline
\endfoot

\hline
\endlastfoot

Existence and Representation & Allow wallets to generate and use self-controlled DIDs and pseudonymous identifiers independently of PID status, with optional binding to official credentials when needed.  \vspace{0.2cm} & Recognize self-generated identifiers for non-regulated uses and limit mandatory PID linkage to cases where high-assurance identification is legally required.  \vspace{0.2cm}\\
Decentralization, Transparency and Autonomy & Support autonomous DID creation, decentralized credential-status checks, and transparent trust infrastructure instead of relying only on PKI-bound verification.  \vspace{0.2cm}& Formally accommodate decentralized trust models, including DLT-based infrastructures, where accountability and public oversight remain verifiable. \vspace{0.2cm}\\
Ownership. Single Source, control and consent & Bind PIDs and attestations to user-controlled DIDs, support secure wallet credential import and export, and make issuer---and holder---signed consented presentations verifiable from the wallet. \vspace{0.2cm}& Treat the wallet as the primary point of consent and require explicit, auditable authorization for credential issuance, sharing, revocation, and reuse. \vspace{0.2cm} \\
Privacy and Minimal Disclosure & Make selective-disclosure credential formats, context-separated PID-bound DIDs, salted claims, and untraceable presentation mechanisms standard. \vspace{0.2cm}& Turn selective disclosure and anti-linkability safeguards into mandatory requirements, and narrowly define when relying parties may request identifying data. \vspace{0.2cm}\\
Verifiability and Authenticity & Support  DID-bound attestations, signed presentation requests, and public status registries so credentials remain verifiable without live issuer dependence. \vspace{0.2cm}& Recognize cryptographic proofs and public witnesses on DLTs as valid evidence of authenticity, reducing exclusive reliance on institution-managed trust lists. \vspace{0.2cm}\\
Security and Protection & Mandate strong audited cryptography, local key control, open implementations, user-auditable logs, and resilient ledger-based integrity mechanisms. \vspace{0.2cm}& Set binding security baselines and liability rules that let users prove misuse, denial, or unauthorized access without depending only on provider records or implicit trust in privacy-preserving compliance. \vspace{0.2cm}\\
Accessibility and Availability & Support local storage and verification of DIDs, VCs, and revocation data, plus peer-to-peer access paths that keep services usable during outages. \vspace{0.2cm}& Prevent any single provider or authority from unilaterally blocking access to credential use or verification, especially for essential identity functions. \vspace{0.2cm}\\
Recoverability, Persistence and Interoperability & Support non-custodial key and credentials management, decentralized recovery, long-lived resolvable DIDs, public witnesses, and open DID/VC standards across wallets and borders. \vspace{0.2cm}& Require recovery and migration rights, and decouple continued identity validity from continuous PID or provider validation except where specific high-assurance checks are legally necessary. \vspace{0.2cm}\\
\end{longtable}
\normalsize
\setlength{\tabcolsep}{6pt}
\renewcommand{\arraystretch}{1}

}
\normalsize

\begin{figure}[htbp]

\centering 
    \includegraphics[width=0.8\textwidth]{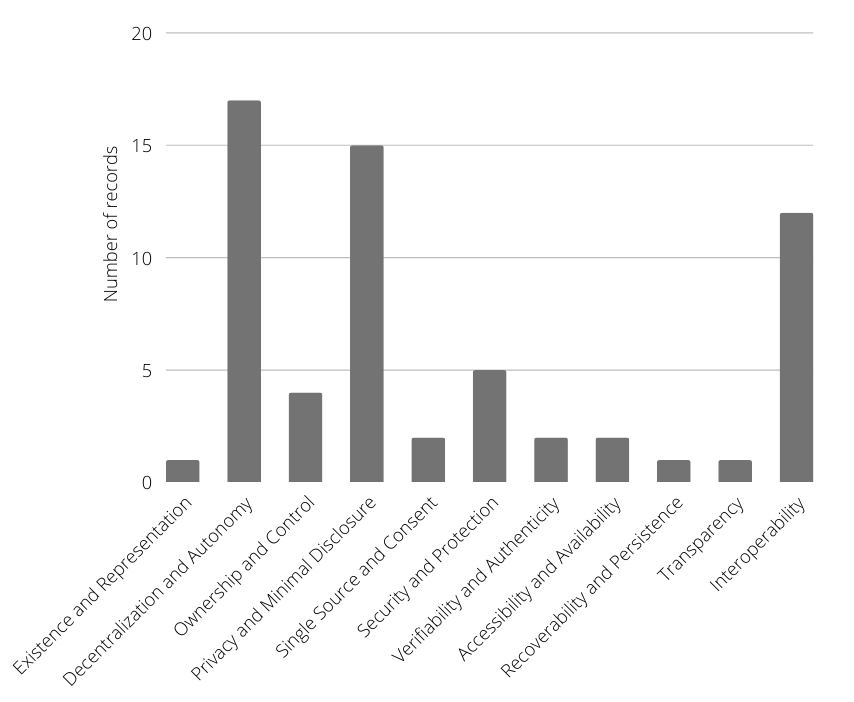}
    \caption{ Graphical representation of research interest on eIDAS 2.0 conformity per each relevant SSI principle }
    \label{Graphical representation of research interest on eIDAS 2.0}
    \end{figure}

The SLR also reveals uneven scholarly attention across SSI aspects, as illustrated in Fig. \ref{Graphical representation of research interest on eIDAS 2.0}, according to the SSI principles defined in \cite{9858139}.

\subsection{Governance and Design Tradeoffs}
 \label{Governance and Design Tradeoffs}
A recurring issue identified in our analysis concerns the legal requirement for EUDIW authentication, which \textcolor{black}{requires, for legally recognized electronic identification,}  the use of a single, valid, government-issued PIDs to access the EUDIW services. From the perspective of SSI maximalists \cite{biedermann2025aggregating, satybaldy2024taxonomy}, the right of digital representation is expected to mirror an individual’s physical existence, with identity wallets envisioned as functional analogues of physical multi-purpose wallets. Under this view, a PID can only correspond to a physical government-issued identity card and cannot fully represent an individual’s digital presence. Moreover, this requirement places PID providers in a position where privacy and minimal disclosure cannot be preserved, due to inherent linkability to pseudonyms and anonymous credentials in the same wallet \cite{alvarez2025privacy, kutylowski2025pseudonymization}. 

\textcolor{black}{Another line of discussion arises from the consequences of defining electronic identification explicitly as the verification of a government-issued PID. Such a definition positions other credentials as technically verifiable attestations that nonetheless lack legal recognition, creating ambiguity around the role of EAAs that are cryptographically verifiable and even regulated, yet insufficient to establish legal identity. This ambiguity becomes particularly relevant in contexts where accountability is central: in practice, liability is typically attributed to the issuer under civil law, while self-issued assertions offer limited recourse with respect to the holder, as responsibility falls on them. Does this hinder the adoption of SSI if such credentials are not formally recognized? Not necessarily. As long as EAAs emonstrate sufficient maturity and security, especially when supported by decentralized infrastructures such as blockchain networks.  Indeed, eIDAS 2.0 already regulates many of the foundational components required for SSI models, particularly attribute-based credentials, thereby opening a pathway toward the broader acceptance of more flexible and expressive forms of identification and authentication as they mature.}

The ARF design of the EUDIW, based on  wallet instances and their WUAs, offers accessibility and conventional security assurances for users who may not be able to manage a fully self-administered system \cite{bukhari2024defining, saaida2023digital}. At the same time, it centralizes accountability and makes users dependent on specific providers, whose empowerment has consistently been associated with privilege abuse and privacy violations \cite{canbay2022privacy, barth2022understanding}. The most realistic direction is a model in which regulated actors retain liability functions, while wallets remain open, interoperable, and portable, and users retain stronger control over identifiers, credentials, and disclosure choices.\\
The above tradeoffs are, in fact, reconcilable. The EUDIW could be implemented as an open-source framework of standards and protocols, ensuring interoperability through exportable and importable credentials. Such a design would allow reliance on wallet providers---similar to custodial wallets in Web3 \cite{seymour2024custodial}---while also supporting SSI-style autonomous identifiers and credentials.

\subsection{GDPR Constraints on DLT-Based EUDIW Models}
\label{GDPR Constraints on DLT}
\textcolor{black}{A decentralized-model of the EUDIW remains constrained by the current legal treatment of DLTs: as the GDPR does not explicitly govern blockchain systems. Immutability and limited access controls of Blockchain technologies can be seen as conflicting properties \cite{suripeddi2021blockchain, rotondi2019distributed}, while the literature also showcases that their spread and acceptance  by businesses and government projects might indicate otherwise \cite{suripeddi2021blockchain, lyons2018blockchain}. eIDAS 2.0 does recognize DLTs' potential for long-term preservation of data, only in compliance with GDPR. Existing tensions between GDPR and blockchain technology can be summarized as: (1) the ambiguity in defining data controllers and processors, where roles vary across private and public blockchains, with unclear responsibilities for participants such as validators and developers; (2) challenges around anonymization, as the absence of standardized techniques and the risk of re-identification mean that most on-chain data remains subject to GDPR and at risk of disclosure; and (3) conflicts with data subject rights, particularly due to blockchain immutability, which complicates requirements such as data erasure and territorial compliance.}

\subsection{Limits and Open Challenges of PET Integration}
\label{pets limits}
To improve compliance, proposed solutions focus on minimizing and obfuscating personal data stored on-chain while preserving data subject rights. Privacy-preserving techniques such as  Homomorphic Encryption (HE), Trusted Execution Environments (TEEs), Secure Multi-Party Computation (SMPC), and ZK proofs enable validation and computation without disclosing personal data \cite{10.1145/3633477}. These PET families are not equally mature for EUDI deployment. As summarized in \ref{pets}, current selective-disclosure stacks such as mdoc/ISO~18013-5 and, to a slightly lesser extent, SD-JWT-based flows are materially closer to deployment readiness than broader ZKP, HE, or SMPC-based approaches. 

While most PETs are effective in protecting data during processing and use, they are less frequently applied to long-term storage or archival contexts. This gap may stem from limited familiarity among practitioners in those domains, as well as from performance constraints. As scalability is inherently tied to performance, PETs must balance strong security guarantees with computational efficiency.\\
HE is widely explored in privacy-sensitive domains such as biometric authentication, but it faces several limitations. Many HE schemes support only restricted classes of computations and struggle with more complex operations. They may also be exposed to risks such as decoding or side-channel attacks, while their design limits the ability to monitor intermediate computations---allowing potential malicious activities to be invisible. Additionally, HE introduces significant computational overhead, increased ciphertext size, and latency (e.g., authentication operations may take several seconds), which complicates practical deployment \cite{s23010014}. Similarly, SMPC enables collaborative computation without data disclosure but introduces communication overhead and reliance on multiple participating parties, which can affect scalability and robustness \cite{10180048}. ZK proofs provide strong privacy-preserving verification mechanisms applicable to both centralized and distributed systems; however, they remain computationally demanding, face interoperability challenges, and often require trusted setup assumptions \cite{gupta2025zero}.

Confidential computing approaches based on TEEs offer a more  and in practice more deployable alternative by enabling secure processing of sensitive data, including identity-related workloads or custodial wallet operations \cite{sabanic2025confidential, anasuri2023confidential, sitouah2026enabling}. Nevertheless, TEEs reintroduce trust dependencies on specific hardware vendors, which may undermine self-sovereignty objectives—particularly in scenarios where the TEE supply chain becomes concentrated or monopolized.
In contrast, reversible encryption remains risky due to potential key exposure, while irreversible hashing occupies a gray area because of transaction traceability, necessitating safeguards such as one-time values when used in notarial contexts \cite{lyons2018blockchain}. Complementary approaches address data subject rights by aggregating personal information~---~using mechanisms such as Merkle trees or private blockchains that anchor aggregated data to public chains---and by enabling rectification through mutable encryption and controlled update mechanisms, including Chameleon hashes, which allow data modification without compromising integrity \cite{neri2015does, lima2018blockchain, sitouah2025blockchain}.
Additional challenges arise from the fact that certain PETs are optional in the regulation rather than mandatory, limiting the effective application of pseudonymization and anonymization for providers and privileged users. Moreover, insufficient attention is given to critical functionalities such as recovery mechanisms \cite{bochnia2024long, woo2016improving, singh2021private}, untraceability of credential usage \cite{sitouah2024untraceable, 11260514, podda2025impact}, and transparency with verifiable accountability \cite{kutylowski2025pseudonymization}.

\section{ Conclusion}
 \label{chap6}
This article aims to raise concerns and highlight the importance of identity regulation for privacy protection, particularly with regard to personal authentication data and the control of identity attributes. In the context of EU citizens, eIDAS legislation plays a key role in defining the quality of their digital experience, leading to extensive scrutiny and debate on the envisaged e-identity model, which is expected to be widely adopted. 

We conducted an in-depth study of the eIDAS regulation and comprehensively analyzed the SSI principles in the literature. The study focuses on the evaluation of SSI and blockchain technologies with respect to the GDPR, eIDAS 2.0 and its ARF. In addition, we present a use case for complying with both the SSI and eIDAS principles, as well as exhaustive details on how each SSI property can be preserved without violating eIDAS 2.0 or the GDPR.

\textcolor{black}{In summary, eIDAS 2.0 incorporates attribute-based credentials--- largely compatible with SSI concepts---within a structured framework, centered on government-issued PIDs, EUDI Wallets/WUA providers. It further encourages the use of certain PETs, while leaving their adoption to providers discretion. As a result, assessing compatibility with SSI becomes more nuanced: each SSI property finds partial alignment within the framework, yet specific design and regulatory choices may limit full conformity.}
However, it does open the possibility for practical integration, beginning with the legal recognition of DLTs, which were previously not accepted for authoritative records. Which emphasize that researchers should focus on improving the understanding of decentralized applications, consensus types, and the privacy inherited from application-level encryption built on top of blockchain networks. 
\subsection{Contribution of this Study}
Our objective was to address the following research questions:
\begin{enumerate}
\item[RQ1]: To what extent are SSI principles compatible with the eIDAS 2.0 Architecture Reference Framework?
\item[RQ2]: What technical or regulatory modifications could improve alignment with GDPR and ARF requirements while preserving SSI principles?
\item[RQ3]: Which challenges arise specifically from misalignments between DLTs and existing regulations?
\end{enumerate}
 
Concerning RQ1, our analysis identifies several areas in which SSI aligns with the regulation, as well as others that require further adaptation. Centralization remains a dominant paradigm in the development of the EUDIW, \textcolor{black}{architecturally and partially } constraining SSI principles related to sovereignty and user autonomy. This approach is also strongly authoritative, as government-issued attestations (PIDs) play a foundational role in the proposed EUDIW architecture. Consequently, users remain dependent on PID providers, EUDIW providers, certification bodies, and centralized, web-based PKI endpoints to verify the validity of their identities and credentials (including revocation, update, etc).

\noindent Even though the regulation does not explicitly oppose DLTs, their use is currently restricted, primarily to backup storage functionalities. Our findings further indicate that particular SSI principles: decentralization, privacy, and interoperability, are more consistently addressed in the literature. In contrast, principles such as recoverability, persistence, and existence remain comparatively underexplored. Thus, achieving a mature and integration-ready SSI ecosystem requires rigorous and comprehensive research across all SSI principles.

Regarding RQ2, our analysis yields two main implications. First, it suggests the integration of additional features that could \textcolor{black}{more fully} incorporate selected SSI properties into the existing model, while acknowledging that \textcolor{black}{the architectural choice and the extent to which specific technologies are mandated or merely recommended under eiDAS 2.0 as operationalized via the EUDIW ARF, may constrain this potential and limit full conformity}. 
Second, it argues for the reconsideration of certain regulatory aspects, including the rigid reliance on government-issued identifiers (PIDs), centralized trust infrastructures, and optional privacy safeguards, advocating instead for more flexible, interoperable, and user-centric models. This includes the adoption of alternative security measures (PETs) and cryptographic accountability mechanisms to minimize disclosure risks, strengthen transparency, and enhance user control, while also considering the anonymity-related implications, including the potential risk of facilitating organized cybercrime.

At last, with regards to RQ3, we identified three primary tensions between blockchain and the GDPR from the literature: the lack of explicit distinction between data controllers and processors, challenges with anonymization in standard transactions, and the immutability of on-chain data. Roles within a blockchain-based SSI system should be clearly defined in accordance with eIDAS 2.0 and the GDPR. In combination with a European consensus protocol, this definition should ensure that unauthorized users are restricted in their issuance capabilities, while authorities and their representatives retain recognized issuance and credential revocation powers without compromising user privacy or control over personal data. Furthermore, data minimization and obfuscation techniques should be applied to on-chain information, and PETs should enable validation and verification while preserving user privacy.

Overall, this work highlights the need for continued research from academia and industry to better understand eIDAS 2.0 and support informed regulatory development around SSI and decentralization. \textcolor{black}{Furthermore, the shift toward SSI-compatible mechanisms, such as verifiable credentials in place of traditional OAuth-based approaches, presents a significant opportunity for research to advance their maturity, robustness, and trustworthiness, thereby supporting a potential future progression toward SSI from both technological and regulatory perspectives.  }

\subsection{Future Work}
 
At the current stage, SSI and DLTs in general are neither fully legally recognized nor widely adopted as normalized solutions---socially embraced. This situation opens new research avenues, including privacy-enhancing technologies (PETs), Confidential Computing (CC), and bridging solutions \cite{biedermann2025aggregating, lepore2024aligning, moser2025privacy, Stamoulis2025}. PETs like such as zero-knowledge proofs, proofs of signature \cite{0xparc2021zkecdsa}, and Merkle tree–based cryptographic schemes \cite{sitouah2024untraceable, azeem2023urs} can enhance the privacy and security of existing identity systems, including the EUDIW. Confidential computing, which relies on Trusted Execution Environments (TEEs) \cite{b1}, provides isolated and secure computation environments that can support cloud-based EUDIW solutions for users lacking compliant devices, and may also enable the deployment of SSI-compliant wallets in the cloud.
\vspace{0.5cm}

\newpage
\appendix
\section{Abbreviation}
\noindent\textbf{SSI} Self-Sovereign Identity
\\\textbf{ARF} Architecture and Reference Framework
\\\textbf{eIDAS} 
\\\textbf{EC} European Comission
\\\textbf{EU} European Union
\\\textbf{GDPR} General Data Protection Regulation
\\\textbf{eID} electronic IDentification  
\\\textbf{eIDAS}  electronic IDentification, Authentication,
and Trust Services
\\ \textbf{EUDI} European Digital Identity
\\\textbf{IDMS} Identity Management System 
\\\textbf{Dapps} Decentralized applications 
\\\textbf{DLT} Decentralized Ledger Technology
\\\textbf{PET}  Privacy Enhancing Techniques 
\\\textbf{OIC} OpenID Connect
\\ \textbf{SAML} Security Assertion Markup Language  
\\\textbf{DID} Decentralized Identifiers
\\\textbf{CMP} Consent Management Platforms 
\\\textbf{VC} Verifiable Credentials
\\\textbf{LoA} Level of Assurance
\\\textbf{EUDIW} European Digital Identity Wallet
\\\textbf{CAB} Conformity assessment body 
\\\textbf{NAB} National Accreditation Body 
 \\\textbf{EUDIWP} EUDIW Provider
 \\\textbf{WUA} Wallet Unit Attestation 
 \\\textbf{(Q)-TSP} (Qualified) Trust Service Provider  
 \\\textbf{(Q)-EAA} (Qualified) Electronic Attestation of Attributes  
 \\\textbf{PuB-EAA} Public Body Authentic Source Electronic Attestation of Attributes 
 \\\textbf{PID} Personal Identification Data
 \\\textbf{(Q)-ES } Qualified) Electronic signatures 
 \\\textbf{PET} Privacy Enhancing Technologies
 \\\textbf{ZKP} Zero Knowledge Proof
 \\\textbf{AC} Anonymous Credential
 \\\textbf{MNO} Mobile Network Operator
 \\\textbf{TOTP} Time-based One-time Password 
 \\\textbf{ OID4VC} OpenID for Verifiable Credentials
  \\\textbf{ OID4VCI} OpenID for Verifiable Credentials Issuance
 \\\textbf{OID4VP} OpenID for Verifiable Presentations
  \\\textbf{WSCD} Wallet Secure Cryptographic Device
 \\\textbf{WSCA} Wallet Secure Cryptographic Application
 \\\textbf{PKI} Public Key Infrastructure
  \\\textbf{LDAP} Lightweight Directory Access Protocol
 \\\textbf{HE} Homomorphic Encryption 
 \\\textbf{SMPC} Secure Multi-Party Computation 
 \\\textbf{CSR}  Certificate Signing Requests 
  \\\textbf{HMAC} Hash-based Message Authentication Code 
 \\\textbf{CRL} Credential Revocation List
 \\\textbf{CMP} Consent Management Platform
 \\\textbf{AS} authentic sources  
 \\\textbf{SD-VC}  Selective Disclosure Verifiable Credentials    
 \\\textbf{SD-VP} Selective Disclosure Verifiable Presentations  
 \\\textbf{EAP} Extensible Authentication Protocol  
 \\\textbf{eMRTD}   electronic Machine-Readable Travel Document
   \\\textbf{CSA} CyberSecurity Act 
 \\\textbf{ETSI} European Telecommunications Standards Institute
 \\\textbf{EBSI} European Blockchain Services Infrastructure

\newpage
 \section{Assessment Rubric and Traceability Matrix}
 \label{app:assessment-rubric-traceability}

This appendix consolidates the coding rubric and the traceability matrix used for the \textcolor{black}{four-level} classification. In the "eIDAS 2.0 / ARF basis" column only explicit relevant numbered references   cited in Section~\ref{Analysis: SSI Property} are included, thereby strengthening the normative assessment.

\begin{table}[ht]
\centering
\footnotesize
\setlength{\tabcolsep}{4pt}
\renewcommand{\arraystretch}{1.08}
\caption{\textcolor{black}{Four-level classification used for the compatibility assessment.}}
\begin{tabular}{@{}>{\raggedright\arraybackslash}p{4.2cm}>{\raggedright\arraybackslash}p{10.0cm}@{}}\\
\hline 
\textbf{Label} &  \textbf{Decision rule} \\
\hline
Compatible & The property can be exercised in practice under eIDAS 2.0 and the ARF without a dependency that defeats its SSI purpose. \vspace{0.2cm}\\ 

\textcolor{black}{Partially-compatible: Architecturally limited} & \textcolor{black}{The property can be supported but constrained due structural and architectural governance patterns. } \vspace{0.2cm}\\

\textcolor{black}{partially-compatible: Optional coverage} & \textcolor{black}{The property can be supported only under limited implementation choices that are not mandated upon providers or considered optional services. }\vspace{0.2cm}\\
Non-compatible & The legal or architectural design prevents the property from being meaningfully exercised in practice. \\
\hline
\end{tabular}

\label{tab:annex-rubric}
\end{table}

\setlength{\LTleft}{0pt}
\setlength{\LTright}{0pt}

\setlength{\tabcolsep}{3pt}

\scriptsize
\begin{longtable}{@{}>{\raggedright\arraybackslash}p{2.35cm}>{\raggedright\arraybackslash}p{3.00cm}>{\raggedright\arraybackslash}p{2.15cm}>{\raggedright\arraybackslash}p{3.15cm}>{\raggedright\arraybackslash}p{4.85cm}@{}}
\caption{Traceability matrix for the clustered SSI properties assessed in Section~\ref{Analysis: SSI Property}.}
\label{tab:annex-traceability} \\
\hline
\textbf{SSI property} & \textbf{Operational question} & \textbf{Classification} & \textbf{eIDAS 2.0 / ARF basis} & \textbf{Reasoning and conditionality} \\
\hline
\endfirsthead

\hline
\textbf{SSI property} & \textbf{Operational question} & \textbf{Classification} & \textbf{eIDAS 2.0 / ARF basis} & \textbf{Reasoning and conditionality} \\
\hline
\endhead

\hline
\endfoot

\hline
\endlastfoot

Existence and Representation &
Can a user obtain and use an EUDIW identity without first depending on a unique state-issued PID and provider approval of the wallet instance? &
\textcolor{black}{Partially-compatible: Architecturally limited} 

 &
Art.~7(d); Art.~5e(1--2); Art.~5a(15); Recital.~15, 19, 29 &
A valid EUDIW use depends on a PID issued by a designated authority and on continued provider trust in the wallet instance via the WUA. Alternative access to services may exist outside the wallet, but self-representation within the EUDIW is \textcolor{black}{partially restricted by government-issued} identity and provider validation. \\
\hline

Decentralization, Transparency and Autonomy &
Can users verify and use credentials with transparent and autonomous operation rather than provider-controlled trust anchors and remote secure-element arrangements? & \textcolor{black}{Partially-compatible: Architecturally limited} 
 &
Recital.~13, 19, 33, 36, 56, 68; Art.~5a(4) &
Self-claimed assertions, pseudonyms, and source-code disclosure support some autonomy and transparency, but credential verification still depends on provider-mediated trust anchors and, in practice, centralized PKI or remote WSCD/WSCA arrangements. Which also affects autonomy, as data flows through the wallet can be observed or mediated by the provider; pseudonymous use may be subject to tracking, and transparency is further constrained as wallet providers may deploy closed-source software. \\
\hline

Ownership, Single source, control and consent &
Does the user remain the single source and effective controller of identity data, or do PID activation, wallet-instance revocation, and authentic-source checks control its use? &
Non-compatible &
Recital.~2, 3, 4 5, 7, 61; Art.~5a(4.a), 5a(14), 5a(5.e) &
Although the regulation repeatedly promises user control, the EUDIW still depends on PID activation, provider-controlled wallet-instance validity, and authentic-source verification for (Q)EAAs. Consent is informed and mediated, but not equivalent to full user ownership or a true single source of identity data. \\
\hline
Privacy and Minimal Disclosure &
Can users disclose only the necessary data as a guaranteed capability, without dependence on issuer or RP disclosure settings or traceable status checks? &
\textcolor{black}{Partially-compatible: Optional coverage and Architecturally limited} &
Recital.~15, 59; Art.~5a(5.a.iii) &
Selective disclosure is strongly supported in principle, with the ARF use of suitable credential formats. But certification does not require (enforce) it , effective minimization depends on issuer disclosure policies and RP acceptance, and untraceability of credential usage and status is not addressed. \\
\hline

Verifiability and Authenticity &
Can wallet presentations and attestations be authenticated and verified reliably, including pseudonymous use where the law permits it? &
\textcolor{black}{Partially-compatible: Architecturally limited}   &
Art.~1(c); Art.~5a(4.a); Art.~5b(5,9); Art.~5f(2); Recital.~18, 22, 61,  &
This pertains presentation flows, public RP registration, certificate-based legitimacy checks, signatures or seals, and pseudonymous presentation/traceability.
\\
\hline

Security and Protection &
Does the framework provide high technical security and user protection without leaving proof of misuse dependent on provider-controlled evidence? &
\textcolor{black}{Partially-compatible:  Architecturally limited}  &
Recital.~15, 28; Art.~13(1); Art.~21(1); Art.~46a(4) &
High-assurance on-boarding, certified schemes, conformity assessment, supervision, liability, WSCD/WSCA use, and wallet-instance revocation give strong institutional security. SSI-style protection remains nonetheless incomplete, as reliance on providers and authorities for proof of misuse or access violations prevents user-verifiable accountability, unlike in transparent decentralized networks. \\
\hline

Accessibility and Availability &
Can users and RPs access and verify identity data continuously without issuer-managed verification platforms or similar intermediary bottlenecks? &
\textcolor{black}{Partially-compatible: Optional coverage and Architecturally limited}  &
Art.~12b; Recital.~15 &
Accessibility is framed as a right, but certificate, attestation, and wallet checks still run through issuer or provider platforms. The analysis also highlights missing scalability provisions and the lack of a decentralized high-availability approach. \\
\hline

Recoverability, Persistence and Interoperability &
Can credentials be recovered, remain usable over time, and function across borders without continuous PID or provider validation? &
\textcolor{black}{Partially-compatible: Optional coverage} &
Art.~45j(1.c); Art. 45j(1.c); Recital.~8, 11, 15, 19, 34, 36, 42; ARF ch.~4.4.3; C(2024)/8495-8496-8498-8507-8516); C(2024)/2979 &
While interoperability is presented as a major goal backed by harmonized rules and formats, recoverability is largely left outside the ARF, and persistence is possible for non-qualified credentials provided decentralized architecture is supported. \\
\hline

\end{longtable}

\newpage
\section{PET maturity overview}
\label{pets}

\setlength{\LTleft}{0pt}
\setlength{\LTright}{0pt}
\setlength{\tabcolsep}{3pt}
\renewcommand{\arraystretch}{1.08}

\scriptsize
\begin{longtable}{@{}>{\raggedright\arraybackslash}p{2.70cm}>{\centering\arraybackslash}p{1.40cm}>{\raggedright\arraybackslash}p{4.80cm}>{\raggedright\arraybackslash}p{2.30cm}>{\raggedright\arraybackslash}p{3.85cm}@{}}
\caption{Indicative technical maturity of PET families most relevant to EUDI Wallet privacy.}
\label{tab:pet-maturity-comparison} \\
\hline
\textbf{PET family} & \textbf{Maturity} & \textbf{Evidence} & \textbf{EUDI role} & \textbf{Main challenges} \\
\hline
\endfirsthead

\hline
\textbf{PET family} & \textbf{Maturity} & \textbf{Evidence} & \textbf{EUDI role} & \textbf{Main challenges} \\
\hline
\endhead

\hline
\endfoot

\hline
\endlastfoot

mdoc / OID4VC / OID4VP

& High
& ISO/IEC~18013-5 is a published standard; OID4VP~1.0 and HAIP~1.0 are final; OIDF reported 98\% passing on 44 OID4VP/HAIP pairs; the Commission reports pilots with over 550 companies and public authorities \cite{iso18013-5,oid4vp-final,haip-final,oidf-interop-2025,ec-eudi-2026}.
& Current wallet baseline EUDI
& Selective disclosure reduces data exposure, but correlation across repeated use and status checks remains a live privacy issue. \\

SD-JWT / SD-JWT VC
& Moderate
& SD-JWT is standardized in RFC~9901; OID4VC~1.0, OID4VP~1.0, and HAIP~1.0 are final; public interop results cover SD-JWT issuance and presentation; however, SD-JWT VC remains an active Internet-Draft \cite{rfc9901,oid4vci-final,oid4vp-final,haip-final,oidf-interop-2025,sdjwtvc-draft}.
& Near-term baseline candidate
& The credential-format layer is still draft, and selective disclosure alone does not fully solve unlinkability or cross-service correlation. \\

BBS+ anonymous credentials
& Moderate
& The W3C BBS cryptosuite remains a Candidate Recommendation Draft, with implementation feedback still required, and the BBS signature scheme remains an active IRTF Internet-Draft \cite{vc-di-bbs,bbs-signatures-draft}.
& Emerging candidate
& Unlinkability is stronger, but standards convergence, holder binding, and status or revocation handling remain unsettled for EUDI deployment. \\

General ZK proofs for predicates and composite proofs
& Low
& ARF~v2.8.0 keeps ZKPs in scope through Topic~53, and the Wallet Instance lifecycle means operational or valid instances can still perform some actions; however, no specific ZKP scheme or interoperability profile has been selected for the EUDI baseline \cite{eudi-arf-v280,eudi-topic-g-zkp}.
& Architecturally in scope, not candidate yet
& Architectural scope does not yet amount to a converged EUDI baseline as proof systems, format integration, conformance tooling, and secure-hardware dependencies remain open. \\

Trusted Execution Environments / confidential computing
& Moderate
& Remote attestation has a standardized architecture in RATS, and GlobalPlatform maintains mature TEE implementation specifications used in deployed device ecosystems \cite{rfc9334, globalplatform_tee_tls_2024}.
& Complementary infrastructure control
& This is a pragmatic infrastructure control, but it shifts trust toward hardware vendors, attestation services, and supply-chain governance. \\

Homomorphic Encryption / Secure Multi-Party Computation
& Low
& Homomorphic encryption has an ISO standard, but public EUDI-specific interoperability, conformance, and deployment evidence for HE or SMPC remains absent \cite{iso18033he2019}.
& Limited (specialized backend functions)
& Computational cost, protocol complexity, and integration burden make these techniques not suitable for mainstream EUDI issuance and presentation. \\

\end{longtable}

\noindent\footnotesize\textit{Note:} Technical maturity is qualitative and evidence-led. \textit{High} means a finalized or published core standard plus public interoperability or conformance evidence plus public deployment or pilot evidence. \textit{Moderate} means some standards maturity and implementation evidence exists, but at least one of those three maturity signals remains incomplete. \textit{Low} means the PET is mainly architectural or research-stage for EUDI, with no public EUDI-grade interoperability or deployment evidence.

\normalsize

\normalsize
\newpage
 \bibliographystyle{elsarticle-harv}
 
\bibliography{ExportedItems}

\end{document}